\documentclass[11pt,a4paper]{article}
\pdfoutput=1
\usepackage{jcappub}

\usepackage{amsmath,amssymb,amsfonts}
\usepackage{graphicx}

\graphicspath{{figs/}}

\newcommand{\figref}[1]{figure~\ref{#1}}
\newcommand{\Figref}[1]{Figure~\ref{#1}}

\newcommand{\sech}{\mathrm{sech}}

\def\laplacian{Laplacian}
\def\ie{i.e.}
\def\etal{et al.}

\begin{document}

\title{Cosmic Bubble and Domain Wall Instabilities I: Parametric Amplification of Linear Fluctuations}

\author[a,b]{Jonathan Braden,}
\author[a]{J.\ Richard Bond}
\author[c]{and Laura Mersini-Houghton}

\affiliation[a]{CITA, University of Toronto, 60 St. George Street, Toronto, ON, M5S 3H8, Canada}
\affiliation[b]{Department of Physics, University of Toronto, 60 St. George Street, Toronto, ON, M5S 3H8, Canada}
\affiliation[c]{Department of Physics and Astronomy, University of North Carolina-Chapel Hill, NC 27599-3255, USA}

\emailAdd{jbraden@cita.utoronto.ca}
\emailAdd{bond@cita.utoronto.ca}
\emailAdd{mersini@physics.unc.edu}

\abstract{This is the first paper in a series where we study collisions of nucleated bubbles taking into account the effects of small initial (quantum) fluctuations in a fully 3+1-dimensional setting.
In this paper, we consider the evolution of linear fluctuations around highly symmetric though inhomogeneous backgrounds.
In particular, we demonstrate that a large degree of asymmetry develops over time from tiny initial fluctuations superposed upon planar and SO(2,1) symmetric backgrounds.
These fluctuations are inevitable consequences of zero-point vacuum oscillations, so excluding them by enforcing a high degree of spatial symmetry is inconsistent in a quantum treatment.
To simplify the analysis we consider the limit of two colliding planar walls, with mode functions for the fluctuations characterized by the wavenumber transverse to the collision direction and a longitudinal shape along the collision direction $x$, which we solve for. In the linear regime, the fluctuations obey a linear wave equation with a time- \emph{and} space-dependent mass $m_{eff}(x,t)$.
In situations where the walls collide multiple times,  $m_{eff}$ oscillates in time.
We use Floquet theory to study the evolution of the fluctuations and generalize the calculations familiar from the preheating literature to the case with many coupled degrees of freedom. The inhomogeneous case has bands of unstable transverse wavenumbers $k_\perp$ whose corresponding mode functions grow exponentially. By examining the detailed spatial structure of the mode functions in $x$, we identify both broad and narrow parametric resonance generalizations of the homogeneous $m_{eff}(t)$ case of preheating. The unstable $k_\perp$ modes are longitudinally localized, yet can be described as quasiparticles in the Bogoliubov sense.
We define an effective occupation number and show they are created in bursts for the case of well-defined collisions in the background. The transverse-longitudinal coupling accompanying nonlinearity radically breaks this localized particle description, with nonseparable 3D modes arising that will be studied in subsequent papers.
}

\date{\today}

\maketitle

\section{Introduction}
\label{sec:linear_intro}
We study the behaviour of linearized fluctuations around colliding scalar field domain walls in situations where the walls possess a high degree of spatial symmetry.
Domain walls arise when discrete symmetries are spontaneously broken and are familiar from the magnetic domains formed in ferromagnets.
In the context of the early universe they can form in both high-temperature and vacuum phase transitions, 
either through self-ordering dynamics following a rapid quench or as the walls of nucleated bubbles during a first-order transition.
However, if the walls are stable, their energy density falls only as the square of the expansion factor, strongly constraining the allowed outcome of  these phase transitions.
Domain walls and similar objects such as D-branes are also a common ingredient in early universe model building.
Examples include braneworld cosmologies in which our observable dimensions are confined to either a lower-dimensional brane or a domain wall embedded in a higher dimensional space~\cite{Randall:1999ee,Rubakov:1983bb,Langlois:2002bb,Martin:2003yh,Bucher:2001it},
inflationary cosmologies including stacks of D-branes~\cite{Burgess:2001fx,Dvali:1998pa}, and some cyclic cosmologies~\cite{Khoury:2001wf}.

In a complete theory the domain walls interact with their own dynamics, either inherited from an underlying scalar field theory or intrinsically in the case of D-branes.
When several such walls are present, the evolution may result in collisions.
In some cases, such as the self-ordering  after a quench or a rapid percolating first-order phase transition, the domain walls form a complicated network with interactions and collisions occurring in a wide variety of orientations.
However, in other scenarios the collisions possess a large amount of symmetry, such as planar symmetry or SO(2,1) symmetry.
These highly symmetric configurations may arise from tuning of the initial conditions as in braneworld cosmologies.
In other cases, the dynamics naturally lead to symmetric collisions, although the underlying theory might still require tuning to realize the appropriate limit (e.g., of  planarity). An example of this is tuning to ensure a slow first order phase transition where collisions are infrequent and the bubbles expand to several times their initial radius before colliding.

Here our focus is on the particular case of colliding parallel planar walls formed by the condensate of some scalar field $\phi$.
The qualitative behaviour of the fluctuations around the planar walls also carries over to the case of collisions with an SO(2,1) symmetry.
These two symmetry assumptions --- planar and SO(2,1) --- are widely invoked to study collisions in braneworld scenarios~\cite{Takamizu:2006gm,Bucher:2001it,Omotani:2011un,BlancoPillado:2003hq} and false vacuum decay~\cite{Hawking:1982ga}, respectively.
In both cases, assuming so much symmetry reduces the underlying field equations to a one-dimensional nonlinear wave equation.
This reduction greatly simplifies the problem and has been central to many past studies of domain wall collisions. We use the solutions to these one-dimensional nonlinear wave equations as backgrounds upon which our fluctuations are superposed.

An important difference between the classical and quantum problems, which has been neglected in previous work, is the extent to which the dynamical evolution preserves the initially assumed symmetries.
While the \emph{classical} dynamics may possess exact planar or SO(2,1) symmetry, the quantum fluctuations break the symmetry in any realization, albeit at quite a small level initially.
These fluctuations may be subject to unstable growth as the combined background plus fluctuation system evolves, with the early stages of the instability described by the linearized perturbation equations we solve for in this paper.
If the fluctuations grow they will have considerable backreaction and rescattering effects on the symmetric part of the field, which can significantly modify the overall dynamics. 
A proper treatment of these effects requires studying the nonlinear problem and is the subject of two companion papers~\cite{ref:bbm2} and~\cite{ref:bbm3}.

We restrict our considerations to two different single-field scalar theories possessing domain wall solutions. 
We refer to this  field $\phi$ as the symmetry breaking scalar field. The background spacetime is assumed to be Minkowski throughout.
In addition to the choice of underlying theory, 
the evolution of the fluctuations depends on the particular background around which we expand the fluctuations.
Therefore, we consider a variety of collisions in each potential.
We show that nonplanar fluctuations in $\phi$ can experience exponential instabilities for a broad class of collisions.

Our analysis uses Floquet theory applied to a non-separable PDE.
This generalizes the techniques used in preheating, where the spatial homogeneity of the background results in Floquet theory applied to the ODE for the fluctuations. 
We find generalizations of broad parametric resonance and narrow parametric resonance to the case of fluctuations around a spatially inhomogeneous background.

Although we focus on two specific scalar field models, the dynamical mechanism that leads to the rapid growth of fluctuations is much more general.
As we will explicitly demonstrate, the broad parametric resonance instability 
is essentially particle production in the Bogoliubov sense for fluctuations bound to the walls.
These fluctuations are the transverse generalization of the Goldstone mode arising from the spontaneous breaking of translation invariance by the domain wall.
Therefore, these modes exist for \emph{any} membrane-like structure appearing in a translation invariant theory.
Examples of such membrane-like structures include domain walls in other field theories and D-branes in string theory.
When the two ``branes'' are well separated, the fluctuations are trapped by an effective potential well.
As long as the shape of these wells is modified by the collision we expect similar instabilities to arise regardless of the underlying theory.

The remainder of this paper explores the rich dynamics of linear fluctuations around colliding domain walls.
We first introduce our two models in section~\ref{sec:model} and present the domain wall solutions that each potential supports.
In section~\ref{sec:background} we introduce our decomposition of the field into a background and fluctuations, followed by a review of the background dynamics.
The central analysis in contained in section~\ref{sec:fluctuations}, where we use Floquet theory to understand the dynamics of the fluctuations.
We provide instability charts for the fluctuations and study the mode functions in detail.
We also comment on the applicability of our results to a broader class of theories.
Finally, in section~\ref{sec:bubbles_linear} we briefly comment on the implications for SO(2,1) bubble collisions. We present our conclusions of this linear study in section~\ref{sec:conclusion_lin}.
Some of the more technical details explaining the construction of approximate background solutions are contained in appendix~\ref{sec:collective}.
Details of our numerical methods and convergence tests demonstrating their superb accuracy and convergence properties are in appendix~\ref{sec:numerics}.

\section{Model Lagrangians and Domain Wall Solutions}
\label{sec:model}
In this section, we introduce the two potentials we consider,  review the domain wall solutions each supports and discuss the types of perturbations that exist around these solutions. 
Since we will ultimately work in three spatial dimensions, we also discuss the embedding of lower-dimensional domain wall solutions in three dimensions.

Our first choice of potential is
\begin{equation}
{\rm the \ sine-Gordon \ model:} \  V(\phi) = \Lambda\left(1 - \cos\left(\frac{\phi}{\phi_0}\right)\right) \, , 
  \label{eqn:sg_potential}
\end{equation}
which supports a family of static inhomogeneous solutions --- kinks --- with profiles given by
\begin{equation}
  \phi_{kink}^{SG}(x) = 4\phi_0\tan^{-1}(e^{\sqrt{\Lambda}\phi_0^{-1}(x-x_0)}) + 2\pi n , \qquad n \in \mathbb{Z} \, .
\end{equation}
These solutions interpolate between neighbouring minima of the potential~\eqref{eqn:sg_potential} with $\phi(\infty) = \phi(-\infty) + 2\pi$, and they are the one dimensional version of domain walls.
Here $x_0$ determines the spatial position of the kink and $n$ is an integer determining which minima the kink interpolates between.
There is also a corresponding antikink solution which is obtained by the substitution $(x-x_0) \to -(x-x_0)$.
Kinks moving at a constant velocity can be obtained by Lorentz boosting the static solution.

At linear order in one spatial dimension the only normalizable localized perturbation of the kink is the zero mode corresponding to an infinitesimal translation of the centre of mass
\begin{equation}
  \delta\phi_{\mathrm{trans}} \propto \partial_x\phi_{\mathrm{kink}} \propto \mathrm{sech}\left(\sqrt{\Lambda}\phi_0^{-1}(x-x_0)\right) \, .
\end{equation}
We will later consider planar kink solutions in two or more spatial dimensions.
For planar walls in higher dimensions, 
these localized perturbations give rise to a spectrum of bound state fluctuations with dispersion relationship $\omega =\vert {\bf k}_{\perp} \vert$, where ${\bf k}_{\perp}$ is the 2D wavenumber parallel to the wall.

\begin{figure}
  \centering
  \includegraphics[width=0.6\linewidth]{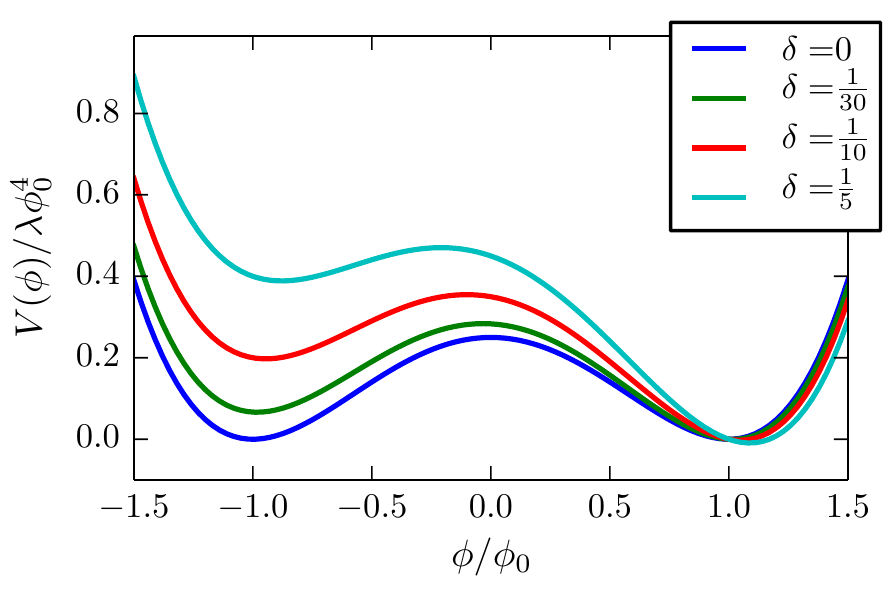}
  \caption[Plots of the double-well potential for several choices of the parameter $\delta$]{Plots of the double-well potential for several choices of the parameter $\delta$ controlling the difference in potential energies between the two wells.  Domain wall solutions interpolate between spatial regions where the field is near the false vacuum at $\phi_{false} \approx -1$ and regions where it is near the true vacuum $\phi_{true} \approx 1$.}
  \label{fig:doubwell}
\end{figure}

Our second example of a potential supporting domain wall solutions is
\begin{equation}
{\rm the \ double-well:} \  V(\phi) = \frac{\lambda}{4}\left( \phi^2 - \phi_0^2 \right)^2 - \delta \lambda\phi_0^3\left(\phi - \phi_0\right) + V_0
  \label{eqn:potential}
\end{equation}
depicted in \figref{fig:doubwell}.
$\delta$ is an adjustable parameter controlling the difference between the false and true vacuum energies and $V_0$ is a constant.\footnote{Unless explicitly indicated, for the remainder of the paper we measure all dimensionful quantities in units of $m_{norm}$, with the exception of the fields measured in units of $\phi_0$ and the potential in units of $m_{norm}^2\phi_0^2$.  The natural mass scale $m_{norm}$ is given by $m_{SG}=\sqrt{\Lambda}\phi_0^{-1}$ for the sine-Gordon potential and $m=\sqrt{\lambda}\phi_0$ for the double-well potential.}
As long as $\delta$ is not too large, this potential supports spatially dependent field configurations in which the field is localized near each of the minima in different regions of space, with the requisite domain-wall structures interpolating between the different regions.

A well known example occurs for $\delta=0$ in one spatial dimension.
In this case, the kink solution located at $x_0$ is given by
\begin{equation}
  \phi_{kink}^{DW}(x) = \phi_0\tanh\left(\frac{\sqrt{\lambda}\phi_0(x-x_o)}{\sqrt{2}}\right)
  \label{eqn:phikink}
\end{equation}
with a corresponding antikink solution again obtained by replacing $(x-x_0) \to -(x-x_0)$.
Moving kinks are again obtained by Lorentz boosting.
Unlike the sine-Gordon model, which posesses a single bound state excitation, 
the one-dimensional double-well kink has two localized normalizable linear perturbations.
There is also a continuum of delocalized radiative modes with frequencies $\omega^2 \ge 2$.
The localized perturbations are often referred to as the translation mode
\begin{equation}
  \delta\phi_0 \propto \partial_x\phi_{kink} \propto \mathrm{sech}^2\left(\frac{x-x_o}{\sqrt{2}}\right)
  \label{eqn:transmode}
\end{equation}
and the shape mode
\begin{equation}
  \delta\phi_1 \propto \cos(\omega t)\sinh\left(\frac{x-x_0}{\sqrt{2}}\right)\mathrm{sech}^2\left(\frac{x-x_0}{\sqrt{2}}\right) \, , \qquad \omega^2=\frac{3}{2} \, .
  \label{eqn:shapemode}
\end{equation}
The translation mode corresponds to a spatial translation of the centre of the kink and is thus analogous to the sine-Gordon zero mode. The shape mode is an internal excitation which can be thought of as an oscillating wall width.

Kink-like solutions continue to exist as we deform the potential by increasing $\delta$.
However, the potential energy difference between the two minima causes the ``wall'' to accelerate, so the kink solutions are no longer time-independent in inertial reference frames.

In three spatial dimensions domain walls become embedded two-dimensional hypersurfaces with some small but finite width.
We consider two cases that possess a high degree of spatial symmetry.
Our main focus is planar walls generated by extending the sine-Gordon and double-well kink solutions to the additional two spatial directions.
We will also study bubbles of ``true vacuum'' nucleating within the false vacuum in the double-well potential, restricting to choices of $\delta$ in the double-well potential for which these false vacuum bubbles are well described by the Coleman-deLuccia (CdL) instanton~\cite{Coleman:1977py,Callan:1977pt,Coleman:1980aw}. In Minkowski space at zero temperature, the most likely initial bubble profile possesses an SO(4) symmetry in Euclidean signature~\cite{Coleman:1977th} with profile determined by
\begin{equation}
  \frac{\partial^2\phi}{\partial r_E^2} + \frac{d}{r_E}\frac{\partial \phi}{\partial r_E} - V'(\phi) = 0
  \label{eqn:bounce_equation}
\end{equation}
where $r_E^2 = {\bf r}^2 + \tau^2$, $\tau$ is the Euclidean time and $d$ is the number of spatial dimensions.
The initial bubble profile is obtained by analytically continuing back to real time.
In the thin wall limit (valid if the initial radius of the bubble is much greater than the thickness of the wall), 
the friction term in~\eqref{eqn:bounce_equation} is dropped and we are left with the same equation as for a domain wall in the corresponding $1+1$-dimensional theory.
In this limit, the initial bubble radius is given by
\begin{subequations}
  \begin{align}
    R_{init} &= \frac{3\sigma}{\Delta\rho} = \frac{\sqrt{2}}{\delta}\, , \\ 
    \sigma &= \int dr [\partial_r\phi(t=0)]^2 = \frac{2\sqrt{2}\phi_0^2 m}{3}
  \end{align}
\end{subequations}
where $\Delta\rho$ is the difference in energy density between the true and false vacuum and $\sigma$ is the surface tension of the wall.
In the final equalities on each line we made use of the specific form of the potential~\eqref{eqn:potential}.

\section{Dynamics of Planar Symmetric Collisions}
\label{sec:background}
We now study (nonplanar) fluctuations around colliding parallel planar domain walls. We treat this case first because the scale associated with the overall radius of the bubbles does not enter the problem, so parallel domain walls constitute a slightly simpler arena in which to illustrate the underlying fluctuation dynamics. In the limit that the bubbles have radii much larger than any other relevant scale in the problem we also expect bubble collisions to be reasonably approximated by two colliding planar walls. 

\subsection{General Formalism}
Our setup consists of a kink starting at $x=-x_{init}$ and moving to the right and an antikink starting at $x=x_{init}$ and moving to the left.
For ease of nomenclature, we refer to this as a $K\bar{K}$ pair. 
We take the collision axis to be the $x$ direction and
split the field as
\begin{equation}
  \phi(x,y,z,t) = \phi_{bg}(x, t) + \delta\phi(x,y,z,t)
  \label{eqn:fld_background}
\end{equation}
where the fluctuations satisfy $\langle\delta\phi(x,y,z,t)\rangle = 0$. The planar symmetry allows us to approximate ensemble averages with averages over the $(y,z)$ plane.
Before solving for the fluctuation field we must find the background solution around which to perturb.
If backreaction and rescattering effects are ignored the background field $\phi_{bg}$ undergoes the same dynamic evolution as in 1+1-dimensions,  namely
\begin{equation}
  \frac{\partial^2\phi_{bg}}{\partial t^2} - \frac{\partial^2\phi_{bg}}{\partial x^2} + V'(\phi_{bg}) = 0 \, ,
  \label{eqn:wall_background}
\end{equation}
with initial conditions given by the $K\bar{K}$ pair.
Meanwhile, the linearized equation for the fluctuations is
\begin{equation}
  \frac{\partial^2\widetilde{\delta\phi}_{\bf k_\perp}}{\partial t^2} - \frac{\partial^2\widetilde{\delta\phi}_{\bf k_\perp}}{\partial x^2} + \left( k_\perp^2 + V''(\phi_{bg}(x,t)) \right)\widetilde{\delta\phi}_{\bf k_\perp} = 0 
  \label{eqn:wall_linearfluc}
\end{equation}
where $\widetilde{\delta\phi}_{\bf k_\perp}(x,t) = \int dydz e^{-i(k_yy+k_zz)}\delta\phi$ is the 2D Fourier transform of $\delta\phi$ in the directions transverse to the collision axis and $k_\perp^2 \equiv k_y^2+k_z^2$.
We refer to these planar symmetry breaking fluctuations as \emph{transverse}.
At this level of approximation, the fluctuation $\delta\phi$ behaves as a free field with a time and $x$-dependent effective mass ($V''(\phi_{bg}(x,t)$) determined independently by the background evolution.

If the system is discretized in the $x$ direction,  it is just a collection of coupled oscillators.
In real space this coupling occurs via our choice of discretization of the \laplacian\ term, 
while in momentum space (along $x$) the oscillators couple via the Fourier transform of $V''(x,t)$.
This makes it evident that, for a given $V''(x,t)$,  the (time-dependent) 'normal' mode oscillations fully describe the system. Of course, the rate at which the discretized system converges to the continuum result depends upon our choice of spatial discretization. Nonetheless, we will show that this approach can be carried out approximately and it provides useful insights into the behaviour of the fluctuations.

In this paper we use a Fourier pseudospectral approximation to discretize $\partial_{xx}$, thus obtaining exponential convergence as we increase the number of grid sites for a fixed box size. Details about our precise numerical procedures as a well as a demonstration of the superb convergence properties of our techniques can be found in appendix~\ref{sec:numerics}.

\subsection{Dynamics of the Planar Background}

We now briefly review the background dynamics for colliding planar domain walls in our two chosen potentials.
These background solutions determine the form of $V''(\phi_{bg}(x,t))$ to use as input in the fluctuation equation~\eqref{eqn:wall_linearfluc}.
The  planar symmetry allows us to reduce the background dynamics to that of a $K\bar{K}$ pair interacting in 1+1-dimensions. 

\subsubsection{sine-Gordon Model}

\begin{figure}[!ht]
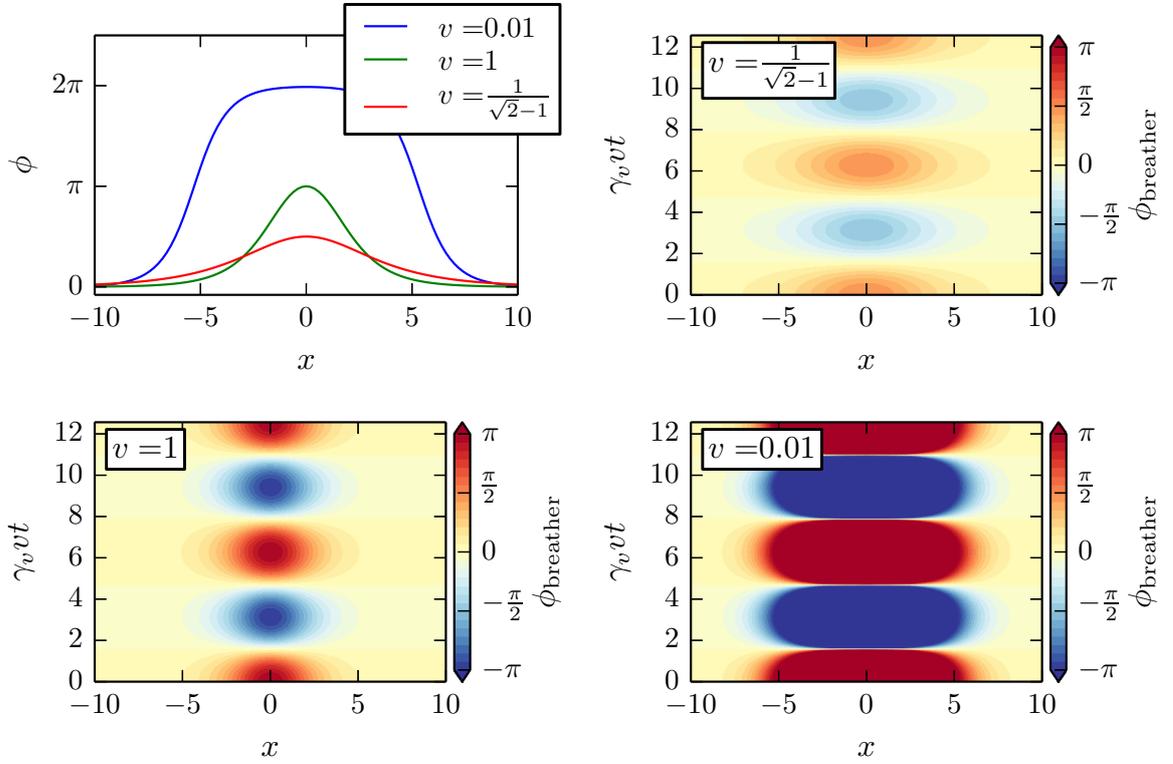

  \begin{tabular}{cc}
    \includegraphics[width=0.48\linewidth]{{{breathers}}} &
    \includegraphics[width=0.48\linewidth]{{{breather_vsqrt2}}} \\
    \includegraphics[width=0.48\linewidth]{{{breather_v1}}} &
    \includegraphics[width=0.48\linewidth]{{{breather_v0.01}}}
  \end{tabular}
  \caption[Three representative cases of breather dynamics]{For the three representative initial $\phi$ profiles shown in the (\emph{top left}) panel,   how $\phi_{breather}$ for the Sine Gordon model evolves is given for $v=(\sqrt{2}-1)^{-1}$ (\emph{top right}), $v=1$ (\emph{bottom left}) and $v=0.01$ (\emph{bottom right}).}
  \label{fig:sg_breathers}
\end{figure}

This model is particularly useful because in one spatial dimension it is integrable and the kink and antikink solutions are true solitons --- they interact with each other while preserving their shapes at early and late times.
More importantly, there exist analytically known periodic breather solutions 
\begin{equation}
  \phi_{breather} = 4\tan^{-1}\left(\frac{\cos(\gamma_v vt)}{v\cosh(\gamma_v x)}\right) \, ,
  \label{eqn:breather}
\end{equation}
where $v>0$ is a free parameter 
determining the properties of the breather solution and $\gamma_v \equiv (1+v^2)^{-1/2}$~\cite{ref:Rajamaran}.
For $v \ll 1$ the kink and antikink pair are well separated at $t=0$ and have initial positions $x_{init} \approx \pm \sqrt{1+v^2}\ln(0.5v)$, whereas
for larger $v$ they are much more tightly bound and the breather is a localized oscillating blob of field with size 
\begin{align}
  r_{breather} &= \sqrt{1+v^2}\cosh^{-1}(1/v\tan(\phi_{edge}/4)) \notag \\
             &\approx -\sqrt{1+v^2}\ln(\phi_{edge} v/8) \ {\rm if} \ \phi_{edge}v \ll 1.  
\end{align}
The edge of the breather is defined to be the point where $\phi_{breather}(t=0,r_{breather}) = \phi_{edge}$. Some representative examples are shown in~\figref{fig:sg_breathers}.

\subsubsection{Double Well Model}

The background evolution in this potential~\eqref{eqn:potential} is  more complicated, but it has been studied by many other authors who detail a rich phenomenology~\cite{McLaughlin:1978,Manton:2004,Drazin:1989,Campbell:1983xu,Manton:1996ex}.
The kinks in this model are solitary waves rather than true solitons, and thus they emit radiation when they interact.  
Combined with excitation of the internal mode during collisions, this means that the motion is no longer exactly periodic.
Because we do not have exact solutions to~\eqref{eqn:wall_background}, we must resort to numerical simulations. 
We use a Gauss Legendre time-integration combined with a Fourier pseudospectral discretization, which allows us to obtain machine precision results for both the spatial discretization and time-evolution.
Details are provided in appendix~\ref{sec:numerics}. 
We restrict our attention to a kink-antikink pair characterized by an initial separation and specified initial velocities. For more general setups,  
the interested reader should consult the previously cited works for the wide range of interesting phenomenology that can arise. 

\begin{figure}[!ht]
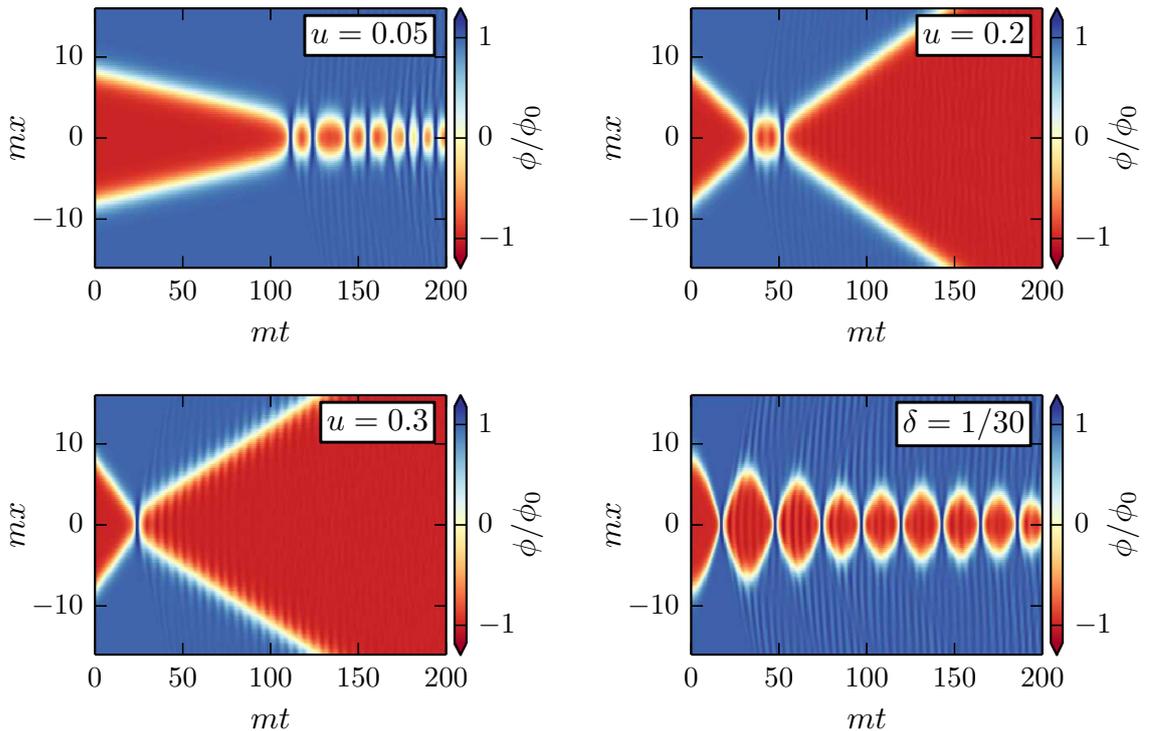

  \centering
  \begin{tabular}{cc}
  \includegraphics[width=0.48\linewidth]{{{1d_collision_v0.05}}}&
  \includegraphics[width=0.48\linewidth]{{{1d_collision_v0.2}}}\\

  \includegraphics[width=0.48\linewidth]{{{1d_collision_v0.3}}}&
  \includegraphics[width=0.48\linewidth]{{{1d_collision_v0_del1o30}}}
  \end{tabular}
  \caption[Sample dynamics of kink-antikink collisions in the double-well potential]{Sample evolutions of the symmetry breaking field $\phi$ in kink-antikink collisions for the double-well potential as a function of time and position along the collision axis.  
Blue corresponds to field values on the true vacuum side of the potential, red to values on the false vacuum side, and white to values near the top of the potential. The spatial coordinates are chosen so that the collision occurs at the origin.  
Three choices of initial velocity in the symmetric well illustrate three different types of behaviour: for $u=0.05$ (\emph{top left}) the $K\bar{K}$ pair capture each other and form a long-lived bound state rather than immediately annihilating. For $u=0.2$ the kinks are in an escape resonance (\emph{top right}), and a collision above the critical escape velocity ($u=0.3$) appears in the bottom left. The bottom-right shows the behaviour for an asymmetric double well with $\delta = 1/30$ and the $K\bar{K}$ pair starting from rest at a separation $d_{sep}=16m^{-1}$, where $m$ is the effective mass. In all cases, the oscillation of the internal shape mode is visible as an oscillating wall width.}
  \label{fig:kink_collision}
\end{figure}

To simplify our analysis we work in the centre of mass frame.
We take the initial kink and antikink speeds $u$ and separations $d_{sep}$ as free parameters.
Some illustrative examples of the dynamics are shown in~\figref{fig:kink_collision}.

In the symmetric well the attractive force between the kink and antikink decreases exponentially at large separations, and as a result unbound motions with the kink escaping back to infinity are possible.
For low initial velocities ($u \lesssim 0.15$), the $K$ and $\bar{K}$ always capture each other after colliding and do not escape back to infinity --- see \figref{fig:kink_collision}, top left, for an example with $u = 0.05$. Rather than immediately annihilating into radiation, the kinks bounce off each other several times and then settle into a long-lived oscillatory blob known as an oscillon (here living in one dimension).
During each oscillation, some energy is radiated away, so this localized state eventually decays.
However, the rate of energy loss is slow and the oscillons can persist for thousands of oscillations (or more) before finally disappearing. 

As the incident velocity increases, the kinks enter a ``resonant escape'' regime characterized by bands of incident velocities in which the two kinks eventually escape back to infinity.
The escape bands are separated by bands in which the $K\bar{K}$ pair trap each other and form an oscillon as in the low velocity limit (see e.g.~\cite{Campbell:1983xu}).
Within each escape band, the number of bounces the walls undergo before escaping back to infinity is a very complicated function of the incident velocity.
This seemingly strange behaviour is usually attributed to the shape mode.
At each collision, nonlinear interactions transfer some of the kinks' translational kinetic energy into a homogeneous excitation of the shape mode or vice versa.
The direction of energy transfer depends on the oscillation phases of the two shape modes (one on the kink and one on the antikink).
As a result, the kinetic energy of the kinks can decrease in one bounce as the shape mode is excited causing the $K\bar{K}$ pair to become temporarily trapped.
In the subsequent collision some of this stored energy can then be transferred back into overall translational motion, giving the kink and antikink enough translational kinetic energy to escape back to infinity.
An example of this behaviour for $u=0.2$ is illustrated in the upper right panel of~\figref{fig:kink_collision}.   
In the first collision, the $K\bar{K}$ pair loses some energy to radiation and excites internal shape modes.  
At the second collision, energy is transferred from the shape mode back into translational kinetic energy and the kinks escape each other.

A critical velocity exists above which the $K\bar{K}$ pair  always interacts exactly once before escaping back to infinity.
During the collision, some of the energy escapes as radiation.
The remaining energy flows between the shape and translational modes of the kink.
Provided that the shape modes are not initially excited, the outgoing velocities of the two walls are thus always less than their incident velocities. This sort of interaction appears in the bottom left panel of~\figref{fig:kink_collision}.

When we take $\delta > 0$, (\ie\ make the double well asymmetric), the difference in vacuum energies across the kink causes an acceleration toward the false vacuum side.
The kink and antikink experience an approximately constant attractive force at large separations.
As a result, the kinks are no longer able to escape back to infinity and will always undergo multiple collisions while slowly radiating energy.
Eventually, an oscillon forms at the location of the original collision. 
See the bottom right panel of \figref{fig:kink_collision} for an example of kinks interacting in an asymmetric double well.

In all cases discussed above the shape mode is excited by the collisions. This is visible as an oscillating wall thickness, as seen in~\figref{fig:kink_collision}.

\section{Dynamics of Linear Fluctuations}
\label{sec:fluctuations}
In the previous section we briefly reviewed the dynamics of planar symmetric background field solutions.
The key feature for our analysis is the presence of oscillatory behaviour: the walls typically bounce off each other many times or settle into a localized bound state rather than immediately annihilating, 
 and internal vibration modes of the walls are excited in the collisions.
We now study the transverse fluctuations described by~\eqref{eqn:wall_linearfluc} in these types of oscillating backgrounds
Our focus is whether small initial (transverse) perturbations to the background dynamics illustrated in~\figref{fig:sg_breathers} and~\figref{fig:kink_collision} can be amplified to the extent that they become important to the full field dynamics.
The results presented in this section are the main findings of our analysis.
We plot (in)stability charts for linear fluctuations around various choices of planar collision backgrounds for both the sine-Gordon and double-well potentials.
In addition, we study several representative choices of the unstable mode functions, which provides insight into the qualitative structure of the instability charts.
We find inhomogeneous generalizations of both broad and narrow parametric resonance.
Our results demonstrate that the transverse fluctuations in $\phi$ can grow rapidly as a result of collisions.

Amplification can happen in several ways. 
Radiative modes can be excited in the collision region and then subsequently propagate into the bulk, which is the transverse generalization of the outgoing radiation seen in the (1+1)-d simulations above. 
However, these excitations carry energy away from the collision and cannot experience sustained growth.
More interesting is the pumping of fluctuations bound to the kinks (in cases where the walls separate widely) 
or bound in the oscillating effective mass well created by the late-time one-dimensional oscillons.
A similar effect for fermionic degrees of freedom has been studied by several authors~\cite{Saffin:2007ja, Saffin:2007qa, Gibbons:2006ge},
and for an additional field coupled to $\phi$ in~\cite{Takamizu:2004rq}.
We only consider fluctuations in the field $\phi$ itself, which is required in a consistent quantum treatment, 
and we do not appeal to additional fields coupled to $\phi$ in order to obtain growing fluctuations.
As well, unlike the authors of~\cite{Takamizu:2004rq} we include the deformation of the spatial structure in $V''$ and couplings between the bound modes and radiative modes.

Although strictly true only for the case of the breathers, we approximate the oscillatory evolution as periodic.
Because of the periodicity, we can use Floquet theory to quantitatively study instabilities in $\delta\phi$.  
The resulting Floquet modes then provide a time-dependent normal mode decomposition in which the evolution of the system is simple.
The periodic approximation is very accurate for the late-time bound states and excitations of the shape mode in the planar background, so Floquet analysis is well justified for those two cases.
When the walls repeatedly bounce off of each other, we show that the amplification happens in a very short time interval around the time of collision.
The short bursts of growth in the fluctuations remain when the bouncing is not periodic, so again the Floquet analysis provides an approximate account of the full evolution.

In early universe cosmology and nonequilibrium field theory, Floquet analysis is familiar from the theory of preheating~\cite{Kofman:1994rk, Shtanov:1994ce, Kofman:1997yn,Allahverdi:2010xz}.
For these problems the background is spatially homogeneous, which leads to two simplifications.
First, the background evolution is described by a nonlinear ODE instead of a nonlinear PDE, and analytic solutions are known in many interesting cases~\cite{Greene:1997fu,Greene:1998pb,Dufaux:2006ee,Abolhasani:2009nb}.
Second, the equation for the fluctuations is diagonalized by the 3D Fourier transform.
Therefore, instead of involving a large number of coupled oscillators the problem reduces to a collection of decoupled oscillators with periodically changing masses $\ddot{\delta\phi_k} + (k^2 + m_{eff}^2(t))\delta\phi_k = 0$.
The effective mass squared $m_{eff}^2(t)=V''(\phi_{bg}(t))$ depends only on time and $k^2=k_x^2+k_y^2+k_z^2$ is now the square of the full three-dimensional wavenumber.
Solutions to this equation have the well known form $\delta\phi_k(t) \sim P(t)e^{\mu t}$ with the (possibly complex) exponent $\mu$ known as a Floquet exponent.
$P(t)$ satisfies $P(t+T) = P(t)$, where $T$ is the period of the oscillation in $m_{eff}^2$.  
The $\mu$'s depend parametrically on $k$ as well as the functional form of $m_{eff}^2$. 
Typically bands of stable ($\max[Re(\mu_k)]= 0$) and unstable ($\max[Re(\mu_k)] > 0$) modes appear as we vary $k$ while holding the form of $m_{eff}^2$ fixed.\footnote{This condition, rather than $\max[Re(\mu_k)] < 0$, is true for the wave equations we study here.
This follows from the fact that the linear operator $M(t) = \left(\begin{array}{cc} 0 & \mathbb{I}\\ \partial_{xx} - V'' & 0\end{array}\right)$ describing the evolution of $(\delta\phi,\delta\dot{\phi})^T$ has $\mathrm{Tr}(M)=0$.  We must then have $\sum_i \mu_i = 0$ so that the presence of a single negative Lyapunov exponent necessarily creates a positive Lyapunov exponent.  For the homogeneous case this requirement becomes $\max[Re(\mu_k)] = 0$.}
There are a variety of underlying mechanisms that can be responsible for the instability, such as tachyonic resonance, narrow resonance and broad parametric resonance~\cite{Shtanov:1994ce,Kofman:1997yn,Dufaux:2006ee}.

For a spatially dependent effective mass, which is the case we consider, the form of the decoupled solutions generalizes to $\delta\phi = F(x,t)e^{\mu t}$, where the profile function $F(x,t)$ satisfies $F(x,t+T)=F(x,t)$.
This is easily seen by discretizing the $x$ direction and placing the system in a finite box.  
The equations of motion then take the form $\dot{\bf v} = M(t){\bf v}$, with $M(t)$ a periodic matrix and ${\bf v} = \left(\delta\vec{\phi}^T,\delta\dot{\vec{\phi}}^T\right)^T$.  
Such equations fall within the domain of Floquet theory and therefore solutions of the form given above are known to exist (modulo issues of convergence on taking the continuum limit).

To treat the case of spatially inhomogeneous masses, we first discretize the fields $\widetilde{\delta\phi}_{k_\perp}$ on a lattice of $N$ points labelled by index $i$.
We denote the field value at the $i$th lattice site by $\phi_i$ and the corresponding field by the vector $\delta\vec{\phi}$.
Next, we consider $2N$ linearly independent solutions $\left(\delta\vec{\phi}^{(j)}(t),\delta\dot{\vec{\phi}}^{(j)}(t)\right)$ to~\eqref{eqn:wall_linearfluc}.
Using this set of solutions, we form a $2N\mathrm{x}2N$ fundamental matrix solution $F(t)$.
For simplicity, we choose our initial conditions such that the $j$th row of $F(t)$ is given by the solution with initial condition
\begin{equation}
  \delta\phi_i^{(j)}(0)= \begin{cases}
    \delta_{i,j} & \text{for $j\le N$} \\ 
    0 & \text{for $j > N$}
    \end{cases} \, ,
  \qquad
  \delta\dot{\phi}_i^{(j)}(0)= \begin{cases}
    0 & \text{for $j \le N$} \\
    \delta_{i,j} & \text{for $j > N$}
  \end{cases} \, .
\end{equation}
Of course, we can construct our fundamental matrix from any complete set of initial states, and we verified that choosing an orthonormal basis in Fourier space reproduced the results we present below.
Finally, the Floquet exponents are related to the eigenvalues $\Lambda_n$ of $F(0)^{-1}F(T)$ via $\Lambda_n = e^{\mu_n T}$,
with the initial conditions for the mode functions given by $F(0){\bf e}^{(n)}$, where ${\bf e}^{(n)}$ is an eigenvector of $F(0)^{-1}F(T)$ with corresponding eigenvalue $\Lambda_n$.
For each choice of $k_\perp^2$ and effective mass there are $2N$ such exponents, but we focus on the exponent with the largest real part (\ie\ the largest Lyapunov exponent).  
For notational simplicity we refer to this maximal real part of a Floquet exponent as $\mu_{max} \equiv \max\left[Re(\mu)\right]$.  
This method allows us to find any exponentially growing instabilities given some fixed background evolution, but it is completely blind to other more slowly growing instabilities.  
In particular, power law growth results when there are degenerate Floquet exponents, and we might expect this to be common in the continuum limit given that we then have infinitely many oscillators.
However we expect the exponentially growing modes to be the most important dynamically and thus most interesting for our purposes.
After all, we are ultimately interested in the quantum problem where we must integrate over all possible modes.

\subsection{sine-Gordon Potential}
\begin{figure}[ht]
  \includegraphics[width=0.48\linewidth]{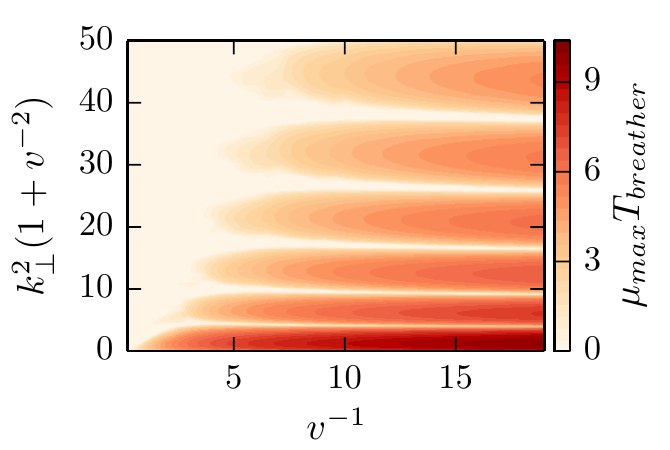}
  \hfill
  \includegraphics[width=0.48\linewidth]{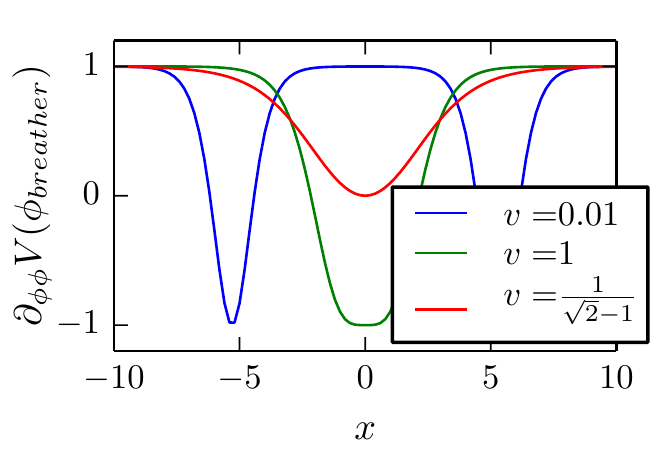}
  \caption[For the sine-Gordon model, the largest real part of a Floquet exponent per period and $V''(\phi_{breather}(x,0))$ for three representative choices of $v$]{\emph{Left}: The largest real part of a Floquet exponent per period of the breather ($\mu_{max}T_{breather}$) for $\ddot{f} - f_{xx} + \left[k_\perp^2 + \cos\left(4\tan^{-1}\left(\frac{\cos(vt/\sqrt{1+v^2})}{v\cosh(x/\sqrt{1+v^2})}\right)\right)\right]f = 0$ as a function of $k_\perp^2(1+v^2)/v^2$ and $v^{-1}$.  The left portion of the chart ($v^{-1} \lesssim 1$) corresponds to the small amplitude breathers while the right portion ($v^{-1} \gtrsim 1$) corresponds to the large amplitude breathers.  \emph{Right}: $V''(\phi_{breather}(x,0))$ for three representative choices of $v$. For $v \ge 1$, there is a localized blob with a single minimum.  We have plotted $v=(\sqrt{2}-1)^{-1}$ and $v=1$ corresponding to the cases when the middle of the breather just reaches $V''(\phi)=0$ and $V''(\phi)=-1$ respectively.  For smaller $v$, the single blob instead develops a pair of minima, with the formation of two distinct wells when $v\ll 1$ corresponding to the well-separated kink and antikink.}
  \label{fig:sg_chart}
\end{figure}

Using the planar symmetric sine-Gordon breather as the background circumvents the problem of finding appropriate approximations to $\phi_{bg}$.
Furthermore, as bound states of a kink-antikink pair, breathers resemble the states arising from collisions in the double-well
and can be used to gain some qualitative insight into the dynamics of the fluctuations for the double-well as well. 
The exact equation for linear fluctuations around the breather is
\begin{equation}
  \frac{\partial^2\widetilde{\delta\phi}_{\bf k_\perp}}{\partial t^2} - \frac{\partial^2\widetilde{\delta\phi}_{\bf k_\perp}}{\partial x^2} + \left[k_\perp^2 + \cos\left(4\tan^{-1}\left(\frac{\cos(\gamma_v vt)}{v\cosh(\gamma_v x)}\right)\right)\right]\widetilde{\delta\phi}_{\bf k_\perp} = 0 \, .
\end{equation}
The initial profile of $V''(x,t)$ for several representative choices of $v$ can be found in the right panel of~\figref{fig:sg_chart}, with the subsequent time evolution for these same three choices in~\figref{fig:v0.01_effmass},~\figref{fig:v1_effmass} and~\figref{fig:vsqrt2_effmass}.

The left panel of~\figref{fig:sg_chart} is an instability chart in which $\mu_{max}T_{breather}$ appears as a function of $k_\perp^2(1+v^2)/v^2$ as we vary $v^{-1}$. 
There are several generic features of note.
First, $\mu_{max} = 0$ when $k_\perp^2=0$ for all values of the parameter $v$ considered.
Since the $k_\perp=0$ modes are part of the planar symmetric system, this is a validation of our numerical procedure since no exponentially growing modes exist in the sine-Gordon model in 1+1-dimension.\footnote{Of course, weaker non-exponential instabilities or transient instabilities are still allowed. 
For example, for a given $v$ we could add a perturbation such that the new field configuration corresponds to a breather with $\tilde{v} \neq v$. Since the period of the breather depends on $v$, the perturbation $\delta\phi = \phi_{breather,\tilde{v}} - \phi_{breather,v}$ will grow initially but will not experience unchecked exponential growth.} 
For $v^{-1} \leq 1$, $V''(x,t)$ has the form of a single well whose depth oscillates with time.
There is a single instability band, the width and strength of which increases monotonically as $v$ decreases.
Once $v^{-1} > 1$ the kink and antikink become less tightly bound and $V''$ develops a pair of local minima separated by a small bump.
An additional instability band then appears on the chart.
As we continue to increase $v^{-1}$ (decrease $v$), the kink and antikink begin to separate from each other during the motion, and the two local minima in $V''$ develop into two well separated wells as seen for $v=0.01$ in the right panel of~\figref{fig:sg_chart}.
As we continue to decrease $v$, additional instability bands appear.
Each of these bands grows quickly with increasing $v^{-1}$ and approaches a nearly constant width, although the strength of the instability (per breather period) within each band continues to increase.
However, $T_{\mathrm{breather}}$ is simultaneously increasing $\sim v^{-1}\sqrt{1+v^2}$ and this results in a decrease of the maximal Floquet exponent if measured in units of $t$. 

We now study the properties of the mode functions for various regimes of $v$.
This gives us insight into the chart's qualitative features and sheds light on the dynamical mechanism responsible for the amplification.

\subsubsection{$v \ll 1$ : Broad Parametric Resonance}
\begin{figure}
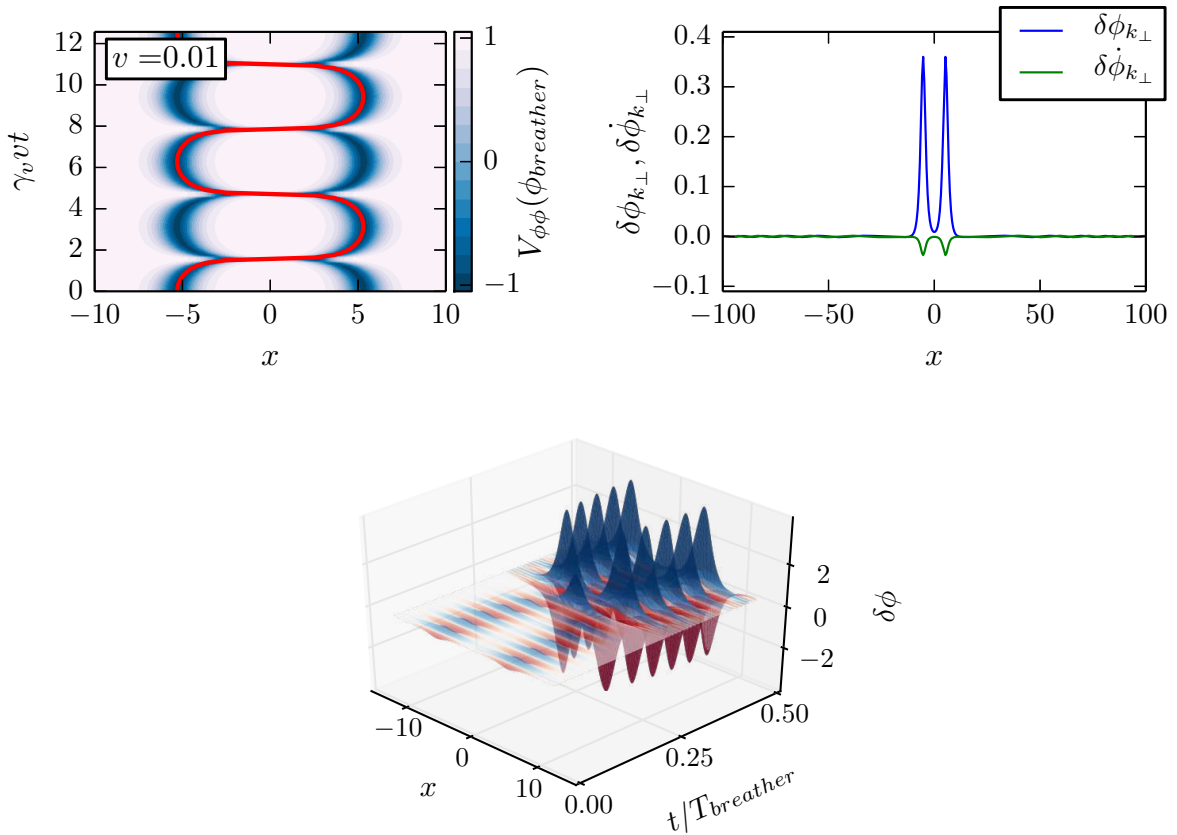

  \begin{center}
  \includegraphics[width=0.48\linewidth]{{{effmass_sg_v0.01}}} 
  \hfill
  \includegraphics[width=0.48\linewidth]{{{sg_mode_initial_v0.01_k0.05}}} \\
  \includegraphics[width=0.6\linewidth]{{{sg_v0.01_k0.05_modefunc3d}}}
  \end{center}
  \caption[For the sine-Gordon model, we show the effective mass and mode function for a large amplitude $v=0.01$ breather and $k_{\perp}^2 = 0.05$]{For the sine-Gordon model, we show the effective mass and mode function for a large amplitude ($v=0.01$) breather and $k_{\perp}^2 = 0.05$.  \emph{Top left:} $V''(\phi_{breather})$ for the given breather background. The instantaneous location of the kink, $x_{kink} = -\frac{\mathrm{sign}(\cos(\gamma vt))}{\gamma}\cosh^{-1}\left(\frac{|\cos(\gamma vt)|}{v}\right)$, is shown as a red line.  \emph{Bottom}: The corresponding mode function, illustrating the constant amplitude oscillations around the kink and antikink and the rapid growth during the short interval when they collide.  Radiation moving away from the collision is also visible.  \emph{Top right}: Initial conditions for the mode function, illustrating both the localization near the kink and antikink, and the extended radiating tail.}
  \label{fig:v0.01_effmass}
\end{figure}

First we consider the $v \ll 1$ limit. 
The breather is effectively a weakly bound kink-antikink pair which repeatedly approach and pass through each other.
The resulting evolution of $V''$ is illustrated in the left panel of~\figref{fig:v0.01_effmass} for the specific case of $v=0.01$.
Due to the reflection symmetry of the potential about any of its minima, 
the period $T_{V''}=\pi\sqrt{1+v^2}/v$ of the effective mass is half the period $T_{breather} =2\pi\sqrt{1+v^2}/v$ of the breather.
For most of the evolution, $V''$ has two distinct wells corresponding to the individual kink and antikink.
In the 1d field theory, each of these wells has a zero mode (the translation mode) associated with it.
When we allow for the transverse fluctuations, the zero mode leads to a continuum of bound excited states with dispersion relation $\omega_{bound} = k_\perp > 0$.
As the breather evolves, the two wells in $V''$ periodically come together and ``annihilate'' each other as seen in the top left panel of~\figref{fig:v0.01_effmass}.
During these brief moments when the two wells have disappeared, the bound states cease to be approximate eigenstates of the system.
{\it The temporary annihilation of the wells allows for the rapid amplification of bound fluctuations.}
Whether or not a particular $k_\perp$ receives coherent contributions to its amplitude at successive annihilations depends on the phase $\omega_{bound}T_{V''} \sim k_\perp\sqrt{1+v^2}/v$ accumulated between collisions. 
In the stability chart this dependence on the accumulated phase manifests as the repeating structure in $k_\perp^2(1+v^2)/v^2$.
Of course, this process is very similar to the familiar case of broad parametric resonance~\cite{Kofman:1997yn}, in which short but large violations of adiabaticity ($|\dot{\Omega}/\Omega^2| \gtrsim 1$) of a harmonic oscillator with a time-dependent frequency $\Omega(t)$ lead to bursts of bound state ``particle production'' in the Bogoliubov sense.
For the familiar homogeneous case the nonadiabaticity is captured by a time-dependent frequency alone, while for the inhomogeneous background we have the additional effect of a strong deformation of the spatial properties of the effective mass.

\begin{figure}
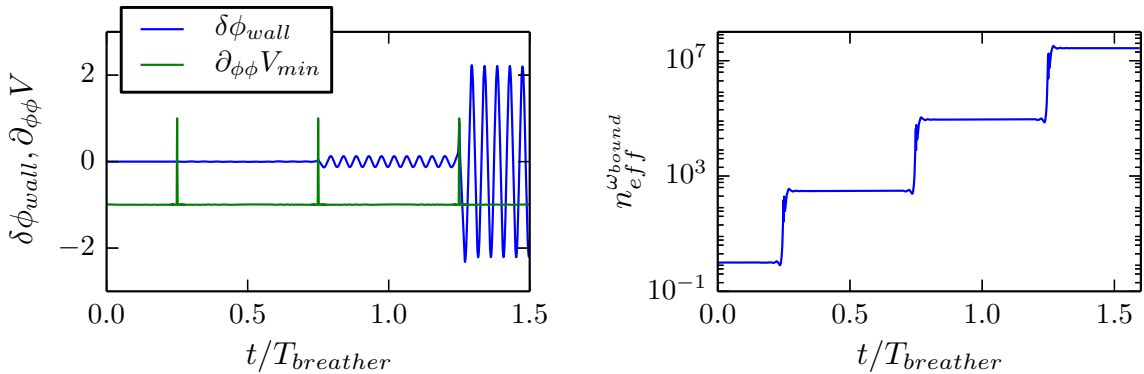

  \begin{center}
  \includegraphics[width=0.48\linewidth]{{{sg_mode_onwall_v0.01_k0.05}}}
  \hfill
  \includegraphics[width=0.48\linewidth]{{{npart_v0.01_k0.05_short}}}
  \end{center}
  \caption[Growth of bound state fluctuations around a $v=0.01$ breather with $k_\perp^2=0.05$]{For the sine-Gordon model, we illustrate the growth of bound state fluctuations around a $v=0.01$ breather with $k_\perp^2=0.05$. \emph{Left}: Mode function evaluated at the instantaneous location of the leftmost kink, defined as the leftmost minimum of $V''$.  For reference we also include the value of $V''$ at the minimum. \emph{Right:} The occupation number of bound state fluctuations as measured by~\eqref{eqn:neff_bound}.  Since we use unnormalized mode functions, we don't have $n_{eff}=0$ initially.}
  \label{fig:v0.01_partnum}
\end{figure}

To illustrate the broad resonance process with a specific example, we plot various aspects of the fastest growing transverse Floquet mode for a breather with $v=0.01$ and $k_\perp^2=0.05$ in~\figref{fig:v0.01_effmass} and~\figref{fig:v0.01_partnum}.
As evident in the bottom panel of~\figref{fig:v0.01_effmass} the mode function is strongly peaked around the locations of the kink and antikink.
The mode function also experiences large jumps around each collision, exactly as expected from our discussion in the previous paragraph.
To further illustrate the rapid amplification of bound fluctuations by the collisions, the left panel of~\figref{fig:v0.01_partnum} shows the value of $\delta\phi$ at the leftmost instaneous minimum of $V''$ as a function of time, along with the value $V''_{min}$ at this minimum.
During the collision, $V''_{min}$ rapidly changes from $-1$ to $1$ and $\delta\phi$ experiences a nearly instantaneous increase in its amplitude.

To quantify the growth of fluctuations, it is useful to introduce the effective particle number 
\begin{equation}
  n_{eff}^{\omega_{bound}} + \frac{1}{2} = \int dx \frac{1}{2k_{\perp}}(k_{\perp}^2\delta\phi_{k_{\perp}}^2+\delta\dot{\phi}_{k_{\perp}}^2)
  \label{eqn:neff_bound}
\end{equation}
which (modulo overall normalization and contributions from the bulk) gives the occupation number of massless transverse fluctuations bound to either the kink or antikink.
For a bound fluctuation on an isolated kink with transverse wavenumber $k_{\perp}$ this quantity is constant.
Therefore, we can use it to track the growth of fluctuations so changes.
From the right panel of~\figref{fig:v0.01_partnum}, we see that between collisions $n_{eff}^{\omega_{bound}}$ is constant and undergoes nearly discontinuous jumps during the short collisions between the kink and antikink.

This analysis confirms our intuitive expectation that the exponential growth results from the production of bound state fluctuations.
However, several additional questions remain that we now address.
First, since the mode function is symmetric about the origin it cannot distinguish between the following processes:
simultaneous amplification of the bound states on both the kink and antikink, amplification of modes bound to the kink (or antikink) only,
and growth of fluctuations in only the left (or right) well.\footnote{\label{fnote:mu_2nd}Actually, we can distinguish the first case from the other two by looking at the second largest eigenvalue.  If pumping only occurred on one kink at a time, then the appropriate linear combinations of only pumping on the kink or only pumping on the antikink would result in two nearly degenerate Floquet exponents. For the sine-Gordon model we have found that within the instability bands the second largest Floquet exponent is always much smaller than the largest one.}
Second, it is unclear whether the stability bands arise primarily from dissipation of fluctuations into the bulk, or primarily from phase interference between bound fluctuations.
Finally, a somewhat surprising feature of the the mode function is the long radiative tail that extends far from the spatially localized breather into the bulk.

\begin{figure}[!t]
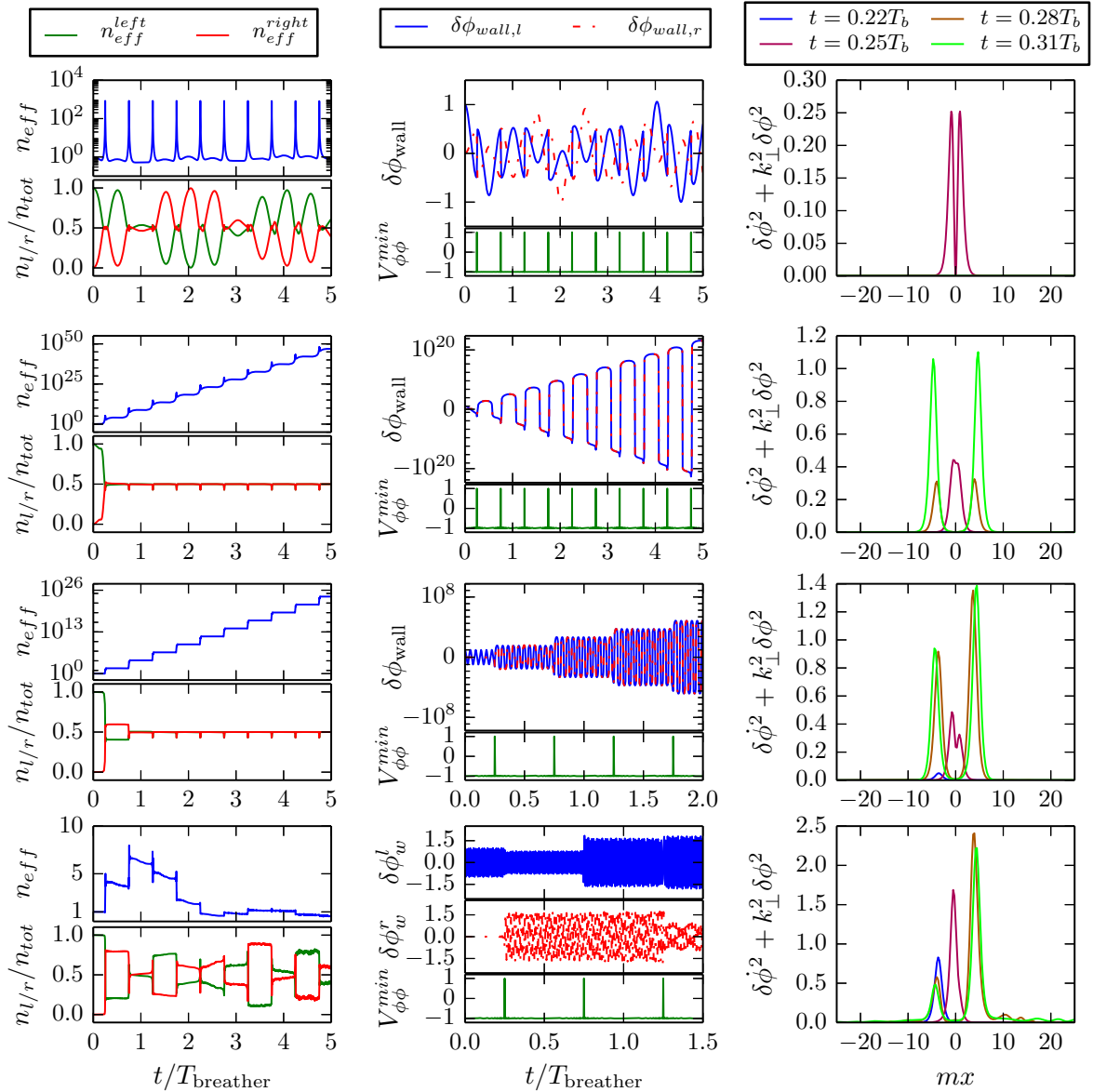

  \includegraphics[width=0.32\linewidth]{{{npart_firststable}}} \hfill
  \includegraphics[width=0.32\linewidth]{{{field_wall_firststable}}} \hfill
  \includegraphics[width=0.32\linewidth]{{{npart_dist_firststable}}} \\
  \includegraphics[width=0.32\linewidth]{{{npart_secondunstable}}} \hfill
  \includegraphics[width=0.32\linewidth]{{{field_wall_secondunstable}}} \hfill
  \includegraphics[width=0.32\linewidth]{{{npart_dist_secondunstable}}} \\
  \includegraphics[width=0.32\linewidth]{{{npart_k0.05}}} \hfill
  \includegraphics[width=0.32\linewidth]{{{field_wall_k0.05}}} \hfill
  \includegraphics[width=0.32\linewidth]{{{npart_dist_k0.05}}} \\
  \includegraphics[width=0.32\linewidth]{{{npart_highstable}}} \hfill
  \includegraphics[width=0.32\linewidth]{{{field_wall_highstable}}} \hfill
  \includegraphics[width=0.32\linewidth]{{{npart_dist_highstable}}} 
  \caption[For the sine-Gordon case, we show the evolution of $\delta\phi$ for initial condition $\delta\phi_{\mathrm{init}} = \sech(x+x_K)$ with $x_K=-\sqrt{1+v^2}\log(v/2)$]{Evolution of $\delta\phi$ for initial condition $\delta\phi_{\mathrm{init}} = \sech(x+x_K)$ with $x_K=-\sqrt{1+v^2}\log(v/2)$.  We used $v=0.01$ and four values of $k_\perp^2$ to illustrate various types of behaviour.  
For each $k_\perp$ we plot: the effective particle number defined in~\eqref{eqn:neff_bound} as well as the fraction of ``particles'' in the left and right half of the simulation (\emph{left}), the value of the field at the instantaneous locations of the two minima in $V''$ (\emph{middle}), and the local ``particle density'' ($\delta\dot{\phi}^2+k_\perp^2\delta\phi^2$) for several times around the first collision (\emph{right}).  In the first row $k_\perp^2=4.006\mathrm{x}10^{-4}$, which is located in the first stability band; in the second row $k_\perp^2=6\mathrm{x}10^{-4}$, which is near the maximum of the second instability band; in the third row $k_\perp^2=0.05$ which is in one of the higher order instability bands; in the fourth row $k_\perp^2=0.83$, which is located in one of the higher order stability bands.  $T_b = T_{breather}$ in the legend for the right panels.
}
  \label{fig:one_kink_fluctuations}
\end{figure}

To gain insight into these issues, we consider fluctuations with the initial condition $\delta\phi_{init} = \sech(x+x_K)$ with $x_K=-\sqrt{1+v^2}\log(v/2)$.
This initial condition corresponds to a bound state fluctuation on the leftmost kink (or antikink) of the breather solution. 
\Figref{fig:one_kink_fluctuations} shows the evolution of this initial state for several choices of $k_\perp^2$.
For each run, we show the evolution of $n_{eff}^{\omega_{bound}}$ defined in~\eqref{eqn:neff_bound}.
Since our setup is no longer symmetric about the origin, we also plot the fractional contributions, $\frac{n_{eff}^{left}}{n_{eff}}$ and $\frac{n_{eff}^{right}}{n_{eff}}$, to the integral from only the left half and right half of the domain respectively.  
For $k_\perp^2=0.83$, a large amount of radiation is produced in the collisions, so we calculated $n_{eff}$ restricted to the interval $|x|<25$ in order to isolate the bound part of the fluctuations.  
The simulation itself had an integration domain of length $L\approx 1174$.
To illustrate the growth of the bound fluctuations, we plot $\delta\phi_{wall}$ evaluated at the instantaneous left and right minima of $V''$.\footnote{The locations at which $V''$ obtains its minimum value is a useful definition of the instantaneous location of the kink and antikink.}
Finally, we show the profile of $\delta\dot{\phi}^2+\delta\phi^2$ at several times around the first collision of the kink and antikink.
This is a useful indication of the growth of fluctuations and in light of~\eqref{eqn:neff_bound} can be interpreted as a local ``particle density''.

In each collision some radiation is released from the collision region.
As $k_\perp$ is increased the fluctuations appear to be less tightly bound to the kink, and the amplitude of radiation produced in the collisions tends to be larger.
This is likely the origin of the decreasing amplitude of $\mu_{max}$ at the centre of the instability bands as we increase $k_\perp$ while holding $v$ constant.
Incidentally, the emitted radiation clarifies the origin of the long radiative tail we found for the mode function in~\figref{fig:v0.01_effmass}.  
Whether or not the mode functions grow, at each collision some radiation is emitted from the collision region.
Consider a pair of subsequent collisions for the case when the mode function is growing exponentially.
Just prior to both collisions, the phase of the mode function will be the either the same (or differ by $\pi$) while the amplitude differs by $e^{\mu T_{V''}}$.
These properties follow from the form $P(x,t)e^{\mu t}$ of the mode function and the step-like nature of the increase in the amplitude.
For simplicity, consider the case when the phases are the same.\footnote{For the case when the mode function picks up a phase factor of $\pi$ between consecutive collisions, we can instead consider three consecutive collisions.  The phase prior to the first and third collisions is then the same and the argument goes through as in the main text.}
Since the fluctuations obey a linear equation, 
the radiation emitted in the second collision is identical to that in the first collision, except it has a larger amplitude by a factor $e^{\mu T_{V''}}$.
Therefore, trains of radiation with the same spatial profile but step-like behaviour in their amplitude will be emitted from the collision.
\Figref{fig:one_kink_fluctuations} also suggests that whether or not a particular value of $k_\perp$ will experience an exponential instability is primarily determined by the phase interference between the bound state fluctuations.
This can occur either because the fluctuations are not excited in any of the collisions (as in the case $k_\perp^2=4.006\mathrm{x}10^{-4}$),
or because the fluctuations are excited in one collision but then de-excited due to phase interference in the subsequent collision (as in the case $k_\perp^2=0.83$).
Finally, as expected from studying the second largest Floquet exponent (see footnote~\ref{fnote:mu_2nd}), we see that for the unstable modes the excitation tends to occur on both the kink and antikink simultaneously.
On the other hand, when the modes are drawn from a stability band the fluctuations on the kink and antikink no longer experience the same degree of excitation or de-excitation at each collision.

\subsubsection{$v \geq 1$ : Narrow Parametric Resonance}
\begin{figure}[ht]
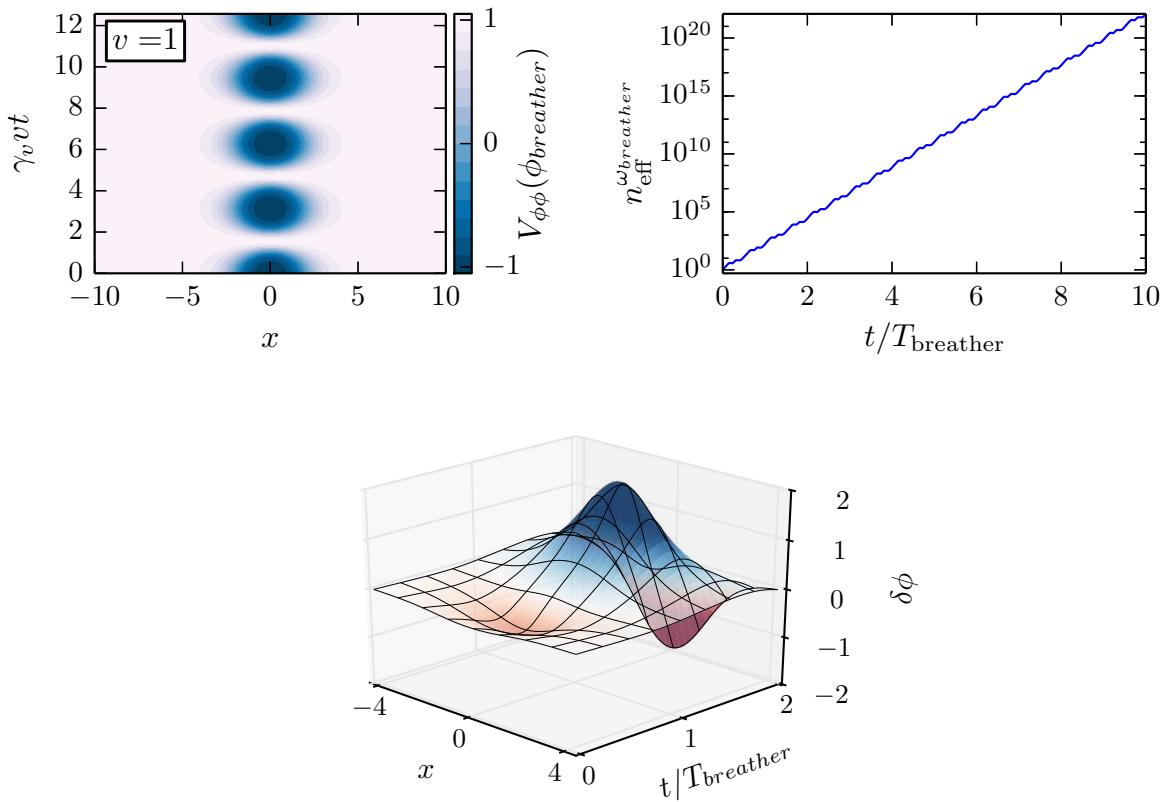

  \begin{center}
  \includegraphics[width=0.48\linewidth]{{{effmass_sg_v1}}}
  \hfill
  \includegraphics[width=0.48\linewidth]{{{npart_v1_k0.35}}} \\
  \includegraphics[width=0.6\linewidth]{{{sg_v1_k0.35_modefunc3d}}}
  \end{center}
  \caption[$V''$ for a breather with $v=1$, fastest growing mode function for $k_\perp^2=0.35$, and $n_{eff}^{(\omega_{breather})}$]{In the top left panel we plot $V''$ for a breather with $v=1$.  In the bottom panel we show the resulting fastest growing mode function for $k_\perp^2=0.35$.  The growth is localized around the location of the breather and is much smoother than the $v\ll 1$ case.  In the top right panel, we illustrate this smooth growth by plotting $n_{eff}^{(\omega_{breather})}$.}
  \label{fig:v1_effmass}
\end{figure}

\begin{figure}[h]
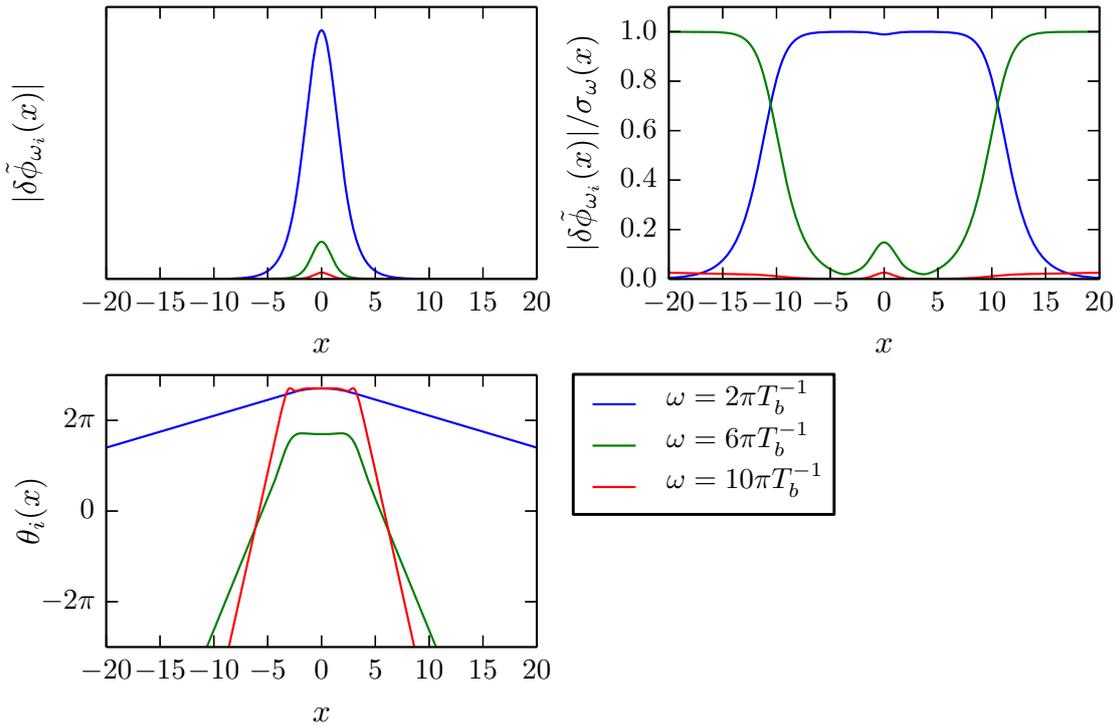

  \includegraphics[width=0.96\linewidth]{{{fourier_v1_multipanel}}}
  \caption[Frequency content of the periodic factor $\delta\phi e^{-\mu t}$ of the fastest growing Floquet mode for the $v=1$ breather with $k_\perp^2=0.35$]{Several aspects of the frequency content of the periodic factor $\delta\phi e^{-\mu t}$ of the fastest growing Floquet mode for the $v=1$ breather with $k_\perp^2=0.35$.
    \emph{Top left:} The amplitude of oscillation $|\delta\tilde{\phi}_{\omega_i}(x)|$.  The overall normalization is arbitrary.
    \emph{Lower left:} The oscillation phase (defined as $\theta_{i}(x) \equiv \tan^{-1}(\mathrm{Im}(\delta\tilde{\phi}_{\omega_i}) / \mathrm{Re}(\delta\tilde{\phi}_{\omega_i})$) for the same frequencies.
    \emph{Top right:} The relative amplitudes of the three largest frequencies normalized to $\sigma_\omega^2(x) \equiv \sum_{i}|\delta\tilde{\phi}_{\omega_i}(x)|^2$.}
      \label{fig:frequency_content_v1}
\end{figure}

We now turn to the case of larger $v$'s, where we no longer have a well-defined kink and antikink at any point in the breather's motion.
We first consider $v=1$ and $k_{\perp}^2=0.35$, which is located near the centre of the instability band.
For this choice of $v$, the middle of the breather just makes it to a maximum of the potential every half oscillation.
Therefore, $V''(x,t)$ has the form of a single oscillating well whose middle oscillates between $-1$ and $1$ and asymptotes to $1$ at $\pm\infty$,
as illustrated in the top left panel of~\figref{fig:v1_effmass}.
The kink and antikink are tightly bound so they never have separate identities.
Therefore, the notion of particles bound to the kink and particles bound to the antikink is ill-defined,
and our previous intuition based on the creation of fluctuations bound to the individual kink and antikink no longer applies.
Instead, we expect the pumping to occur more smoothly and be localized in the region of the breather.
As seen in the bottom panel of~\figref{fig:v1_effmass}, this is indeed the case.
The mode function looks like a smoothly oscillating function with exponentially growing amplitude.
As well, the mode function satisfies $\delta\phi(x,t+T_{V''}) = -\delta\phi(x,t)e^{\mu T_{V''}}$,
so the period of the oscillation is the period of the breather not the period of the effective mass.
One way to see the smooth exponential growth is to consider the quantity
\begin{equation}
  n_{eff}^{(\omega_{breather})} + \frac{1}{2} \equiv \frac{1}{2\omega_{breather}}\int dx\left(\delta\dot{\phi}^2 + \omega_{breather}^2\delta\phi^2 \right) \, .
  \label{eqn:neff_omega}
\end{equation}
$n_{eff}^{(\omega_{breather})}$ is an effective particle number, like~\eqref{eqn:neff_bound} for the $v\ll 1$ breathers, 
but modified to account for the fact that in this case the oscillation frequency of the breather dominates the oscillation frequency of the mode function.
The top right panel of~\figref{fig:v1_effmass} demonstrates the smooth exponential growth of $n_{eff}^{(\omega_{breather})}$ with some small subleading oscillations.
The periodic part of the Floquet modes can be decomposed into an harmonic expansion dominated by the frequency of the breather,
with smaller subleading contributions from higher harmonics, 
\begin{equation}
P(x,t) =   \delta\phi(x,t)e^{-\mu t} = \sum_{\omega_m} |\delta\tilde{\phi}_{\omega_m}(x)|\cos(\omega_mt+\theta_m(x))  \qquad  \omega_m = \frac{2\pi m}{T_{breather}}  \, , 
  \label{eqn:frequency_decomposition}
\end{equation}
as shown in~\figref{fig:frequency_content_v1} .
This dominant frequency justifies our introduction of the quantity $n_{eff}^{(\omega_{breather})}$ in~\eqref{eqn:neff_omega}.
As well, the small oscillations around pure exponential growth of $n_{eff}^{(\omega_{breather})}$ arise from the subleading frequency content.
As expected, the mode function is concentrated at the location of the well created by the breather.
Once we move away from the breather, the spatial phase in each frequency varies linearly.
This linear variation is consistent with the production of radiation in the breather which then travels off to infinity.

\begin{figure}[ht]
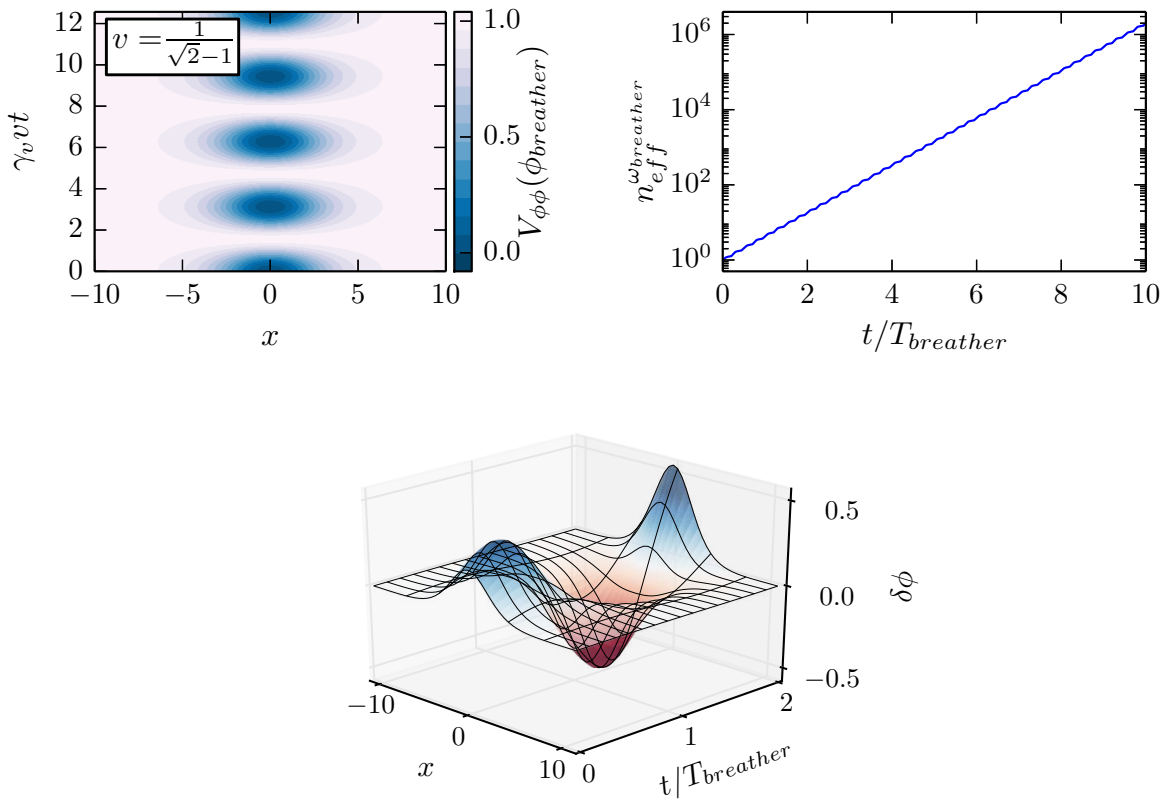

  \begin{center}
  \includegraphics[width=0.48\linewidth]{{{effmass_sg_vsqrt2}}} 
  \hfill
  \includegraphics[width=0.48\linewidth]{{{npart_vsqrt2_k0.2}}}\\
  \includegraphics[width=0.6\linewidth]{{{sg_vsqrt2_k0.2_modefunc3d}}}
  \end{center}
  \caption[Effective mass, mode function and effective particle number for small amplitude breather with $k_\perp^2=0.2$]{For the sine-Gordon model, the effective mass (\emph{top left}), mode function (\emph{bottom}) and effective particle number~\eqref{eqn:neff_omega} (\emph{top right}) for a small amplitude $v=(\sqrt{2}-1)^{-1}$ breather with $k_\perp^2=0.2$.}
  \label{fig:vsqrt2_effmass}
\end{figure}

\begin{figure}[h]
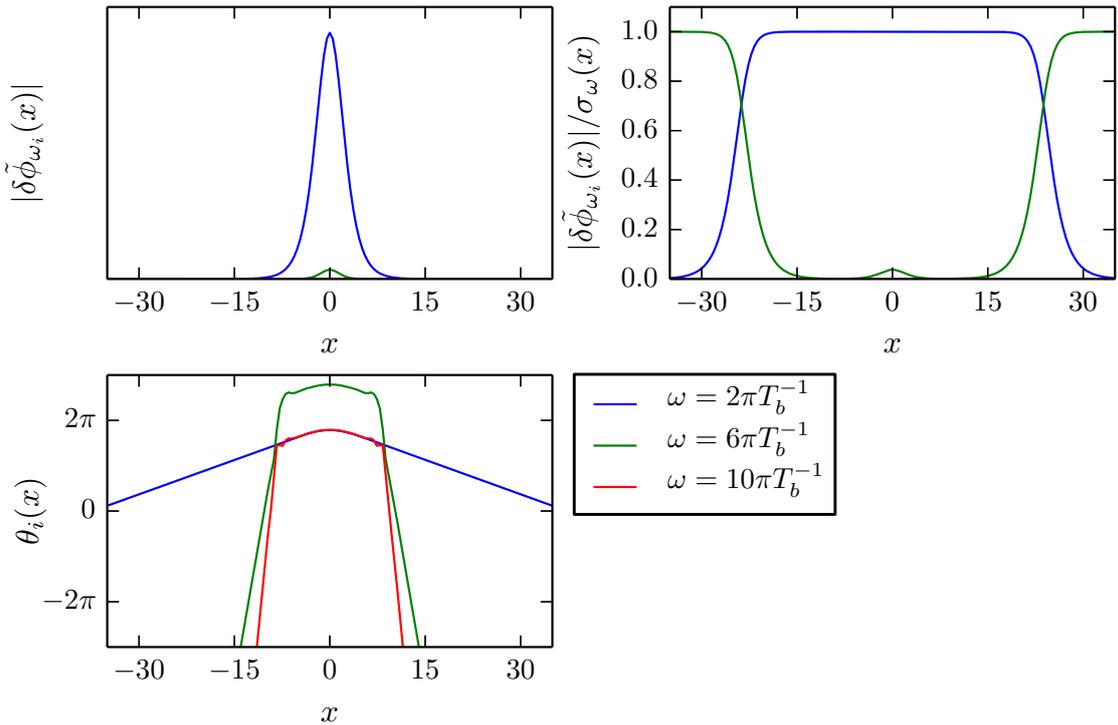

  \includegraphics[width=0.96\linewidth]{{{fourier_vsqrt2_multipanel}}}
  \caption[Frequency content of the fastest growing Floquet mode for $v=(\sqrt{2}-1)^{-1}$ and $k_\perp^2=0.2$]{Frequency content of the fastest growing Floquet mode for $v=(\sqrt{2}-1)^{-1}$ and $k_\perp^2=0.2$.  In the top left panel we show the oscillation amplitude for the three lowest harmonics, while the bottom left panel shows the initial phase of oscillation.  The full solution takes the form $\delta\phi e^{-\mu t} = \sum |\delta\tilde{\phi}_{\omega_i}(x)|\cos(\omega_i t + \theta_i(x))$.  In the top right panel we instead show the relative amplitudes, normalized to the total power $\sigma_w^2(x) \equiv\sum\left|\delta\tilde{\phi}_{\omega_i}(x)\right|^2$ at position $x$.}
  \label{fig:vsqrt2_fourier}
\end{figure}

Finally, we consider the case $v=\frac{1}{\sqrt{2}-1}$ and $k_{\perp}^2=0.2$.
As in the case $v=1$, there is a single oscillating well in $V''$.
However, the maximum excursion of the field is $\pm \frac{\pi}{2}$, so the effective mass for the fluctuations is positive semidefinite.
This is illustrated in the top left panel of~\figref{fig:vsqrt2_effmass}.
The remaining panels of~\figref{fig:vsqrt2_effmass} show that the resulting fastest growing Floquet mode is qualitatively similar to the $v=1$ case.
There is an isolated blob that oscillates with a frequency determined by $\omega_{breather}$ whose amplitude grows smoothly as an exponential.
Looking at the frequency content in~\figref{fig:vsqrt2_fourier}, we see that the mode function again consists of a large amplitude part concentrated
in the potential well of the breather and a smaller amplitude radiative tail propagating away from the breather.
The oscillation frequencies in the two regions are more monochromatic than in the $v=1$ cases.
The dominant frequency of the radiative part differs from the dominant frequency of the bound part.
In both the $v=1$ and $v=\frac{1}{\sqrt{2}-1}$ case, the growth of mode functions is analogous to the case of narrow resonance for a single oscillator.


\subsection{Symmetric Double Well}
We now consider the symmetric double-well.
As demonstrated in~\figref{fig:kink_collision}, for certain choices of the initial kink speed the background solutions can undergo oscillatory motion.
For the case of the asymmetric well, the oscillatory motion is generic due to the attractive force experienced by a well separated $K\bar{K}$ pair.
We will not consider the asymmetric well explicitly, but the behaviour of the fluctuations is qualitatively the same as for the symmetric well.

For the double-well, oscillatory motion of the background solution comes in three forms: repeated collisions of the walls, oscillations of the internal planar shape mode, and evolution of a late-time planar oscillon.
The first of these is analogous to the sine-Gordon breathers with $v \ll 1$, while the last is qualitatively similar to the breathers with $v \geq 1$.
For the repeated collisions and late-time oscillons, the only new feature is that we have $V''(x=0) > V''(x=\infty)$ for part of the evolution.
The oscillations of the planar shape mode are a new feature not present in the sine-Gordon model.
We will study each of these background motions independently, although there are several caveats to this approach.
First of all, the repeated collisions and shape mode oscillations typically occur together in actual background solutions.
In the homogeneous case, the stability chart for a harmonic oscillator with effective mass containing multiple frequencies is not the same as a superposition of the stability charts for each individual frequency~\cite{Braden:2010wd}.
Despite this, from the sine-Gordon analysis above we know that the growth of fluctuations due to the collision will occur in a very short time interval during the actual collision.
The collision also excites the (planar) shape mode in the background, 
which is then free to pump excitations during the much longer intervals while the walls separate from each other.
Since the two amplification effects are temporally separated, we can gain good qualitative understanding by considering the processes in isolation.
The resulting interference from the two mechanisms could then be done using projections on the appropriate Floquet basis (either for the wall collisions or the shape oscillations) just before and after each collision.
Of course, since the exact background is not exactly periodic, the interference effects will tend to get smeared out leading to a smoothing of the instability diagram.
This smoothing is analogous to what happens in the homogeneous case when expansion of the universe is included.

\subsubsection{Bouncing of the Walls : Broad Parametric Resonance}
Keeping the above caveats in mind, we begin with the case where the $K\bar{K}$ pair separate widely from each other between collisions.
Unlike the sine-Gordon breather, we have no analytic solutions for this case.
Also, due to the emission of radiation and the excitation of the shape mode, the background motion is no longer periodic.
In order to create a periodic approximation that is amenable to our Floquet analysis,
we take the background to have the following form
\begin{equation}
  \phi_{bg} = -\tanh\left(\frac{\gamma}{\sqrt{2}}(x-r(t))\right) + \tanh\left(\frac{\gamma}{\sqrt{2}}(x+r(t))\right) - 1 \, ,
  \label{eqn:collective_ansatz}
\end{equation}
where $r(t)$ is a dynamical variable and $\gamma = (1-\dot{r}^2)^{-1/2}$.
This ansatz ignores the production of radiation, excitation of the planar shape mode and any additional deformation of the kink profile due to interactions.
After several further simplifying assumptions, we arrive at the following approximate solution for $r(t)$
\begin{equation}
  r(t) = r_{max} + \frac{1}{2\sqrt{2}}\log\left(\cos^2\left(\frac{\pi t}{T_{walls}}\right) + e^{-2\sqrt{2}(r_{max}-r_{min})}\right) \qquad T_{walls} = \frac{\pi}{2\sqrt{6}}e^{\sqrt{2}r_{max}}
  \label{eqn:separation_evolution}
\end{equation}
where $r_{min}$ is the (minimum) solution to $V_{eff}(r_{min}) = V_{eff}(r_{max})$ with $V_{eff}(r)$ defined in~\eqref{eqn:collective_potential} of appendix~\ref{sec:collective}.
Further details of the construction are provided in appendix~\ref{sec:collective}.
An alternative approach would have been to simply insert a periodic function for $r(t)$ and study the resulting fluctuation behaviour.
However,~\eqref{eqn:separation_evolution} is better justified than a completely ad hoc choice for $r(t)$ because it partially incorporates the full field dynamics.\footnote{As a check, we did insert several other parameterized choices for $r(t)$ ``by hand'' and found banding structure as we varied the parameters.} 

\begin{figure}
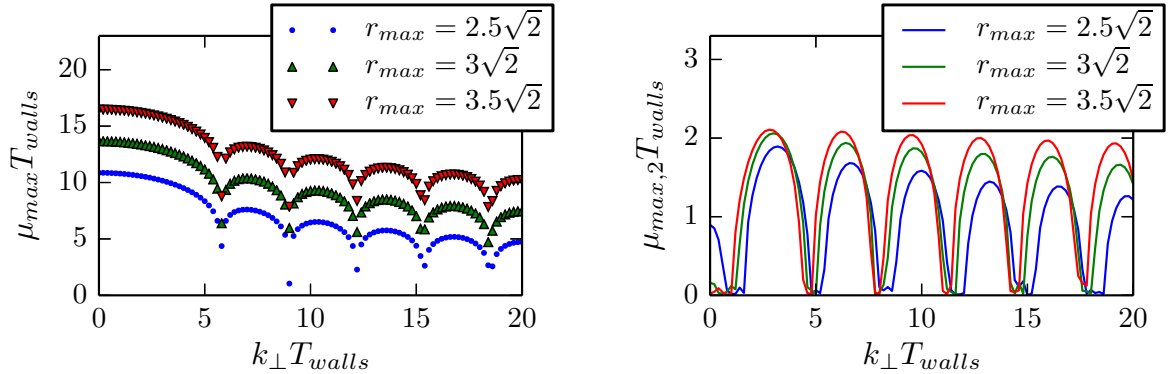

  \centering
  \includegraphics[width=0.48\linewidth]{{{floquet_2walls_max}}}
  \hfill
  \includegraphics[width=0.48\linewidth]{{{floquet_2walls_max2}}}
  \caption[Real parts of Floquet exponents for several choices of initial separation $r_{max}$ of the kink and antikink in the double-well potential]{Real parts of Floquet exponents for several choices of initial separation $r_{max}$ of the kink and antikink in the double-well potential.  In the left panel we show the largest Floquet exponent $\mu_{max}T_{walls}$, while in the right we show the second largest $\mu_{max,2}T_{walls}$.
Unlike the sine-Gordon model, here we have regions where there are two unstable modes rather than just one.  
We see the expected banding structure with bands evenly spaced in $k_\perp T_{walls}$.}
  \label{fig:floquet_twowalls}
\end{figure}

\Figref{fig:floquet_twowalls} shows the resulting Floquet exponents for several choices of $r_{max}$.
As expected, the bands are distributed evenly in the phase accumulated between subsequent collisions by bound fluctuations, $k_\perp T_{walls}$.
There is a very strong instability as $k_\perp \to 0$, which is not unexpected given that our approximation ignores the radiation and planar shape mode that are excited during the collisions.
One interesting new feature is the presence of a second growing mode, which we illustrate by plotting the second largest real part of a Floquet exponent in addition to the largest one. 

\subsubsection{Planar Shape Mode Oscillations : Narrow Parametric Resonance}
Next consider the pumping of fluctuations by oscillations of the planar shape mode.
This is a new effect that is not present in the sine-Gordon model. 
Since the planar shape mode is generically excited (or de-excited) during collisions, 
this source amplification can occur in conjunction with the nonadiabatic production of fluctuations due to the colliding walls described in the previous paragraph.
We parameterize the background motion as
\begin{align}
  \phi_{bg} &= \tanh\left(\xi\right) + A_{shape}\cos(\omega_1 t)\frac{\sinh(\xi)}{\cosh^2(\xi)}, \\ 
  \ {\rm where} \ \xi&\equiv \frac{m(x-x_0)}{\sqrt{2}} \ {\rm and} \  \omega_1^2 \equiv \frac{3}{2}m^2. 
  \label{eqn:background_shape}
\end{align}
As in the case of the repeated collisions described above,~\eqref{eqn:background_shape} is not an exact solution of the 1D field equations.
Nonlinear couplings in the potential will modify the oscillation frequency of the shape mode and also cause it to radiate.
Hence the amplitude will gradually decrease in time, leading to a slowly changing oscillation frequency.
It is even possible to tune the amplitude so that the subsequent evolution leads to the creation of a kink-antikink pair in addition to the kink that was initially present~\cite{Manton:1996ex}.
However, provided the amplitude is not too large, the time-varying amplitude and frequency can be approximated as an adiabatic tracing of modes on the instability chart.
Hence the chart and corresponding mode functions are a good approximation to the fluctuations in the true background.

\begin{figure}[t]
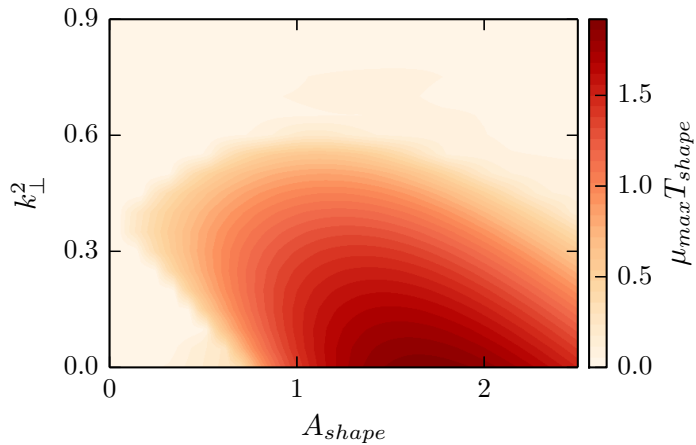

  \begin{center}
  \includegraphics[width=0.6\linewidth]{{{internal_chart_spectral}}}
  \caption[Instability chart for planar oscillation of the shape mode.]{Instability chart for planar oscillations of the shape mode.  
Shown is $\mu_{max}T_{shape}$ for ${\ddot{f} - f_{xx} + \left[\frac{2}{3}(k_{\perp}^2-1) + 2\left(\tanh(x/\sqrt{3}) + A_{\mathrm{shape}}\cos(t)\frac{\sinh(x/\sqrt{3})}{\cosh^2(x/\sqrt{3})}\right)^2 \right]f = 0}$ and various choices of the parameters $k_{\perp}$ and $A_{shape}$.
}
  \label{fig:internal_floquet}
  \end{center}
\end{figure}

Plugging~\eqref{eqn:background_shape} into the equation for fluctuations gives
\begin{equation}
  \delta\ddot{\phi} - \partial_{xx}\delta\phi + \left[k_{\perp}^2 - 1 + 3\left(\tanh\left(\frac{mx}{\sqrt{2}}\right) + A_{shape}\cos\left(\frac{\sqrt{3}mt}{\sqrt{2}}\right)\frac{\sinh(mx/\sqrt{2})}{\cosh^2(mx/\sqrt{2})} \right)^2 \right]\delta\phi = 0 \, , 
\end{equation}
whose Floquet chart is given in~\figref{fig:internal_floquet}.
In the top left panel of~\figref{fig:fourier_internal} we plot $V''(x,t)$ for $A_{shape}=0.5$.
As expected from the interpretation of the shape mode as a perturbation to the width of the kink, $V''$ looks like a well whose width oscillates in time.
There is also some additional oscillating side-lobe structure.
The remaining panels in~\figref{fig:fourier_internal} show the fastest growing Floquet mode, the effective particle number, and the various frequency components of the solution.
These plots are analogous to those for the $v \geq 1$ breathers in~\figref{fig:v1_effmass},~\figref{fig:frequency_content_v1},~\figref{fig:vsqrt2_effmass} and~\figref{fig:vsqrt2_fourier}.
Because the mode function completes only half an oscillation per period of the shape mode, the frequency used in the definition of $n_{eff}^{\omega}$ is $\omega = \pi / T_{shape}$.
The mode function displays more structure than sine-Gordon breather mode functions, in particular in the higher harmonics.
This additional structure is probably due to the additional sidelobe structure of $V''$.
However, the solution is still well described by a single oscillation frequency near the core of the kink, with additional harmonics becoming important as we move into the oscillating sidelobes and then into the radiating regime.
\begin{figure}
  \begin{center}
  \includegraphics[width=0.48\linewidth]{{{internal_a0.5_m2eff}}}
  \hfill
  \includegraphics[width=0.48\linewidth]{{{internal_a0.5_k0.4_npart}}} \\
  \includegraphics[width=0.48\linewidth]{{{internal_a0.5_k0.4_efunc}}} 
  \hfill
  \includegraphics[width=0.48\linewidth]{{{fourier_amp_internal_A0.5_k0.4}}} \\
  \includegraphics[width=0.48\linewidth]{{{fourier_arg_internal_A0.5_k0.4}}}
  \hfill
  \includegraphics[width=0.48\linewidth]{{{fourier_power_internal_A0.5_k0.4}}}
  \end{center}
  \caption[Effective mass, mode function and effective particle number for the unstable mode around the planar shape mode.]{For an unstable mode around an oscillating planar shape mode excitation, the same series of plots are shown as for the Floquet modes of the $v=1$ (\figref{fig:v1_effmass},~\figref{fig:frequency_content_v1}) and $v=(\sqrt{2}-1)^{-1}$ (\figref{fig:vsqrt2_effmass},~\figref{fig:vsqrt2_fourier}) sine-Gordon breathers.
In the definition of $n_{eff}^{\omega}$, we used the frequency $\omega=\pi/T_{shape}$.  As for the sine-Gordon breathers, this mode function consists of a well-defined core region near the location of the kink as well as a much smaller radiative tail.  Due to the additional spatial complexity of $V''$, the mode function displays more spatial and frequency structure than for the breathers.}
  \label{fig:fourier_internal}
\end{figure}

\subsubsection{1D Oscillon Background : Narrow Parametric Resonance}
\begin{figure}
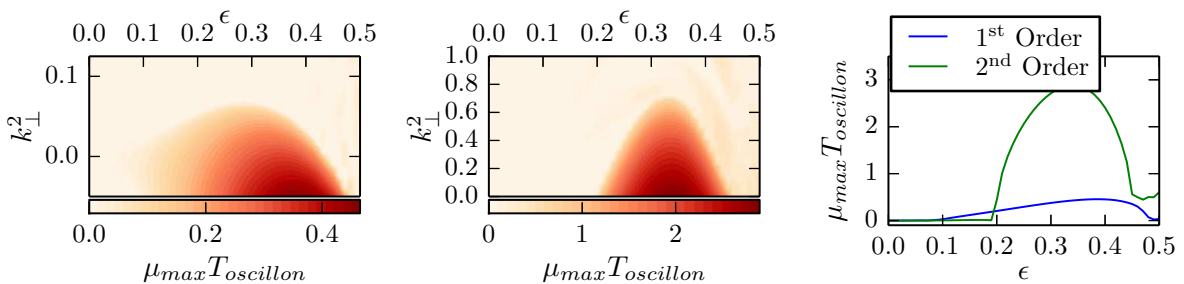

  \centering
  \includegraphics[width=0.32\linewidth]{{{floquet_oscillon}}}
  \hfill
  \includegraphics[width=0.32\linewidth]{{{floquet_oscillon_order2}}}
  \hfill
  \includegraphics[width=0.32\linewidth]{{{floquet_oscillon_kperp0}}}
  \caption[Floquet chart and modes for small amplitude ``breather''-like solution in the double-well potential]{Floquet chart for the small amplitude ``breather''-like solution in the double-well potential, eqn.~\eqref{eqn:breather_doublewell}.  \emph{Left}: Floquet chart is for the leading order in $\epsilon$ solution.  \emph{Middle:} Floquet chart for the second order in $\epsilon$ solution.
\emph{Right:} Instability of the $k_\perp^2=0$ mode as a function of $\epsilon$ for both the leading and second order background solutions.}
  \label{fig:floquet_oscillon}
\end{figure}

Consider fluctuations around the late-time (1-dimensional) oscillon state.
In order to approximate the background motion, we first expand the solution about the true vacuum minimum as $\phi_{bg} = \phi_{min} + \bar{\phi}(x,t)$.  
Up to an additive constant, the potential for $\bar{\phi}$ then takes the form $V(\bar{\phi}) = \frac{\bar{m}^2\bar{\phi}^2}{2} + \frac{\sigma}{3!}\bar{\phi}^3 + \frac{\bar{\lambda}}{4!}\bar{\phi}^4$ with $\bar{m}^2 = 2\lambda\phi_0^2$, $\sigma = 6\lambda\phi_0$ and $\bar{\lambda} = 6\lambda$.
To extract an approximate background solution, we follow~\cite{Fodor:2008es} and perform an asymptotic expansion in some small parameter $\epsilon$ and define new time and space coordinates $u=\sqrt{1-\epsilon^2} \bar{m} t$ and $w=\epsilon \bar{m} x$.
Expanding the solution as $\bar{\phi}(u,w) = \sum_{n=1}^{\infty} \epsilon^n\phi_n(u,w)$ and solving the resulting equations order by order in $\epsilon$, we find
\begin{align}
  \label{eqn:breather_doublewell}
  \notag \bar{\phi}_{\mathrm{osc}} &= (P(x)+g(\sqrt{2} mx))\cos(\sqrt{2}\sqrt{1-\epsilon^2}mt) \\
  \notag      &\qquad+ \frac{3}{2\phi_0}P(x)^2\left(\cos(2\sqrt{2}\sqrt{1-\epsilon^2}mt) -3)\right) + \mathcal{O}(\epsilon^3) \\
  \notag P(x) &= \frac{4\epsilon}{\sqrt{\alpha}}\mathrm{sech}(\sqrt{2}\epsilon mx) \\
   \partial_{ww}g - g +3\alpha g^3 &= 0, \qquad \alpha \equiv \frac{5\sigma^2}{3\bar{m}^4} - \frac{\bar{\lambda}}{\bar{m}^2} \, . 
\end{align}
For the particular form of the double well potential, we have $\alpha = 12/\phi_0^2$.
In the literature, two choices for the function $g$ have been considered.
Fodor \etal~\cite{Fodor:2008es} assumed that no bounded solutions exist and set $g=0$, while Amin~\cite{Amin:2013ika} instead demanded $\phi_2(t=0) = 0$ to set $g(x) = 3P(x)^2/\phi_0$.
We follow Amin when we plot the second order oscillon instability chart.
The resulting equation for the transverse fluctuations is
\begin{equation}
  \frac{\partial^2\delta\phi}{\partial t^2} - \frac{\partial^2\delta\phi}{\partial x^2} + \left[k_2^2 -1 + 3(1+\bar{\phi}_{\mathrm{osc}})^2\right]\delta\phi = 0 \, .
\end{equation}
In~\figref{fig:floquet_oscillon} we show stability charts for the fastest growing mode around the oscillon background~\eqref{eqn:breather_doublewell}.
We plot the charts for both the leading order and second-order in $\epsilon$ approximations to the background.\footnote{The particular choice of oscillon profile is not essential.  We also ran simulations using Gaussian profiles $\phi_{bg} = A\cos(\omega t)e^{-x^2/w^2}$ for various choices of $\omega$ and $w$ taking $A$ as a parameter.
In this case we also found similar banding structure for the Floquet chart.}
The detailed structure of the instability does display some sensitivity to our choice of approximation for the background.
For $k_\perp = 0$ and $\epsilon \lesssim 0.2$, the higher order approximation removes a weak instability that was present in the leading order approximation,
indicating that it is indeed a better approximation to the background at small $\epsilon$.
However, for larger $\epsilon$ the higher-order background is actually more unstable than the leading order approximation.
This is merely a reminder that our approximation is asymptotic rather than convergent.
When we consider $k_\perp \neq 0$, we see that the additional oscillating frequencies in the second-order background lead to several weak instability bands at small $\epsilon$.
For larger $\epsilon$ the main instability band extends for a wider range of $k_\perp$ and has larger Floquet exponents.
This is not surprising since the oscillation amplitude is larger in this approximation and we would thus expect it to drive stronger instabilities.

\subsection{Comments on Collisions of other Membrane-like Objects}
Although we focussed on two specific scalar field models, the dynamical mechanism that leads to rapid growth of the fluctuations is much more general.
In particular, for well-separated walls the explosive growth of fluctuations relied only on the presence of bound fluctuations around each of the kinks
and violation of adiabaticity for these bound states.
Recall that these fluctuations are the transverse generalization of the Goldstone mode (i.e. the translation mode) resulting from the spontaneous breaking of translation invariance by the kink in the one-dimensional theory.
An equivalent Goldstone mode will occur for kinks in any translation invariant theory, and thus these bound fluctuations are ubiquitous.
For kink-antikink type collisions such as those studied here, we typically expect large deformations in the shape of $V''$ during each collision.
Therefore, the parametric amplification of wall-bound fluctuations we have found here should be very common.
We leave for future work the interesting question of kink-kink collisions, where the effective potential wells resulting from each of the kinks do not completely annihilate during collision.

Our results also have relevance to collisions of other membrane-like objects such as D-branes.
When the kink and antikink are well  separated, the transverse translation modes are characterized by a spatially dependent location for the centre of the kink, and are thus well-described by an effective action for a membrane.
If two such membranes are in close proximity to each other, it is natural to expect them to interact.
For the case of D-branes, this interaction is usually described in terms of the excitation and production of string modes.
Since string production is a local process from the viewpoint of a field theory on the brane, we expect that the resulting fluctuations will be inhomogeneous and analogous to the excitation of our transverse modes.
We show in~\cite{ref:bbm2} and~\cite{ref:bbm3} that the inhomogeneity of the growing fluctuations dramatically changes the full three-dimensional dynamics compared to the planar approximation.
The detailed effects of the onset of nonlinearities amongst the fluctuations is dependent on the details of the high-energy theory (in our case a scalar field theory).
However, the linear dynamics described in this paper should be much less sensitive to the UV completion and we expect D-brane collisions to also result in the rapid growth of ``transverse'' fluctuations that eventually require a full nonlinear higher-dimensional treatment.

\section{Comments on Fluctuations Around Colliding Bubbles}
\label{sec:bubbles_linear}
Beginning with the work of Hawking, Moss and Stewart~\cite{Hawking:1982ga},
the two-bubble collision problem has been explored by many authors~\cite{Watkins:1991zt,Kosowsky:1991ua,Chang:2007eq,Aguirre:2008wy,Aguirre:2009ug,Easther:2009ft,Giblin:2010bd,Johnson:2010bn,Gobbetti:2012yq}.
A common feature of these analyses is the assumption of SO(2,1) symmetry for the field profiles.
This is motivated by the SO(4) symmetry of the minimum action bounce solution for a single bubble~\cite{Coleman:1977th},
which translates into an SO(3,1) symmetry for the nucleation and subsequent expansion of the bubble in real time.
The nucleation of a second bubble destroys the boost symmetry along the axis connecting their centres, leaving a residual SO(2,1) symmetry for the 2-bubble solution, 
making of the problem effectively $1+1$-dimensional, 
which greatly eases the numerical challenges and has even led to a general relativistic treatment~\cite{Johnson:2011wt,Wainwright:2013lea,Wainwright:2014pta}.

However this is not the full story.
Quantum fluctuations are present, in fact are responsible for the bubble nucleation in the first place.
We now discuss the linear fluctuation dynamics in the background of the pair of colliding bubbles.
For a discussion of perturbations around a single bubble see~\cite{Adams:1989su,Garriga:1991ts,Garriga:1991tb}.
Results for the fully nonlinear three-dimensional dynamics of bubble collisions are presented in~\cite{ref:bbm3}.

\subsection{Background Dynamics of Highly Symmetric Collisions}

As in the planar wall case, we decompose the field into a symmetric background component and a nonsymmetric fluctuation.
A convenient variable change is
\begin{align}
  t &= s\cosh\chi \notag \\
  x &= x \\
  y &= s\sinh\chi\cos\theta \notag \\
  z &= s\sinh\chi\sin\theta \notag \ ,  
\end{align}
 aligned so  the centres of the two bubbles both lie on the $x$-axis.
The SO(2,1) symmetry is now manifest, as the background depending only on $s$ and $x$.
We separate the field into a symmetric background and symmetry breaking fluctuations $\phi = \phi_{bg}(s,x) + \delta\phi(s,x,\chi,\theta)$.
Ignoring backreaction of the fluctuations, the background solution obeys
\begin{equation}
  \frac{\partial^2\phi_{bg}}{\partial s^2} + \frac{2}{s}\frac{\partial\phi_{bg}}{\partial s} - \frac{\partial^2\phi_{bg}}{\partial x^2} + V'(\phi_{bg}) = 0
  \label{eqn:bubble_bg_I}
\end{equation}
and the linearized fluctuations evolve according to
\begin{equation}
  \frac{\partial^2 (sA_{\ell})}{\partial s^2} - \frac{\partial^2 (sA_{\ell})}{\partial x^2} + 
  \left(\frac{\ell^2}{s^2} +  V''(\phi_{bg}) \right)(sA_{\ell}) = 0 \, .
  \label{eqn:bubble_fluc_I}
\end{equation}
Here we have factored the perturbation into a sum over eigenfunctions, $C_{\ell,n}$ labelled by and integer $n$ and eigenvalues $\ell$ : 
\begin{eqnarray}
\delta\phi = \sum_{\ell,n} A_{\ell}(s,x)C_{\ell,n}(\chi)e^{in\theta} \ ,  \\
    \frac{1}{\sinh(\chi)}\frac{d}{d\chi}\left(\sinh(\chi)\frac{dC_{\ell,n}}{d\chi}\right) = \left(-\ell^2 + \frac{n^2}{\sinh^2(\chi)}\right)C_{\ell,n} \, .
\end{eqnarray}
$\sum_{l,n}$ represents an integral over the continuous part of $\ell$ and a sum over $n$ and any discrete normalizable $\ell$ modes $C_{\ell,n}$.
The curvature of the bubble walls thus influences the fluctuation dynamics in three ways: damping the overall amplitude as $s^{-1}$, redshifting the effective transverse wavenumber squared $\ell^2$ as $s^{-2}$, and modifying the dynamics of $\phi_{bg}$ and by extension $V''(\phi_{bg})$.

\begin{figure}[h]
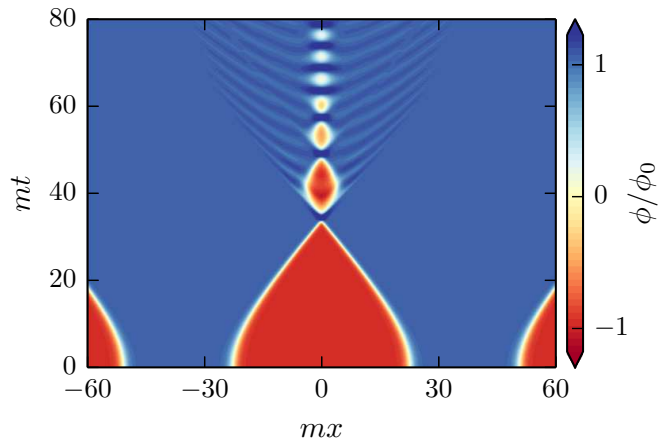

  \centering
  \includegraphics[width=0.6\linewidth]{{{bubble_1d_del0.1}}}
  \caption[Collision of two thin-wall vacuum bubbles in the asymmetric well]{Collision of two thin-wall vacuum bubbles in the asymmetric well~\eqref{eqn:potential} with $\delta=0.1$.  The color coding indicates the value of the scalar field.  Blue indicates it is near the true vacuum minimum, red shows regions where it is near the false vacuum, and the location of the bubble wall is white.  At early times, the acceleration of the walls and corresponding Lorentz contraction is visible.  As in the planar symmetric case, the two walls bounce off of each other multiple times rather than immediately annihilating.  During this process, scalar radiation is emitted from the collision region.}
  \label{fig:bubble_collision_symmetric_I}
\end{figure}

Treatments that assume SO(2,1) symmetry study~\eqref{eqn:bubble_bg_I} and ignore \eqref{eqn:bubble_fluc_I}.
A sample collision between two such bubbles in the asymmetric well with $\delta=\frac{1}{10}$ is shown in~\figref{fig:bubble_collision_symmetric_I}.
As in the case of the kink-antikink collisions, the bubble walls undergo multiple collisions, each time opening up a pocket where the field is localized near the false vacuum minimum.
The bouncing behaviour we observe is characteristic of bubble collisions in double-well potentials with mildly broken $Z_2$ symmetry, and was first noted by Hawking, Moss and Stewart~\cite{Hawking:1982ga}.

Considering the implications of this behaviour for the full $3+1$-dimensional evolution suggests that two instabilities may occur.
The first is the generalization of our previous results to the SO(2,1) symmetric rather than the planar symmetric case.
Given the background evolution depicted in~\figref{fig:bubble_collision_symmetric_I}, we see that~\eqref{eqn:bubble_fluc_I} again describes a field with an oscillating time- and space-dependent mass.
\Figref{fig:bubble_collision_symmetric_I} also reveals the presence of another possible instability.
Because of the SO(2,1) symmetry, each pocket with the field near the false vacuum corresponds to a torus with growing radius centred on the initial collision in the full 3-dimensional evolution.
Roughly, we can think of this as a torus of false vacuum with a thin membrane separating it from the true vacuum on the outside.
The energy difference between the false and true vacuum leads to a pressure acting normal to the local surface of the membrane.
Since this pressure wants to push the membrane into the false vacuum, this will tend to cause small initial ripples on the surface of the torus to grow.
The surface tension of the membrane and the stretching of the torus as it expands tend to counteract this effect,
so that a three-dimensional calculation is required to determine the ultimate fate of these ripples.
As we will see in~\cite{ref:bbm3}, both of these instabilities manifest themselves in the fully 3+1-dimensional problem.

\section{Conclusions}
\label{sec:conclusion_lin}
We studied the dynamics of linear asymmetric fluctuations around highly symmetric collisions between planar domain walls and vacuum bubbles. Parallel planar walls are a common ingredient in cosmological braneworld model building, and SO(2,1) bubble collisions are generally believed to be an accurate description of individual bubble collisions during false vacuum decay.
The results shown here have implications for such cosmological scenarios, as well as being of intrinsic theoretical interest.

Fluctuations are generically present in the field that forms the domain wall and therefore must be included in a quantum treatment of the problem.
However, nearly all past studies of planar wall and vacuum bubble collisions dynamics have used symmetry to reduce the problem to a 1+1-dimensional PDE, thus excluding the fluctuations a priori.
Assessing the validity of this drastic reduction in the effective number of dimensions requires a more sophisticated treatment of the problem, 
and this paper provides such a treatment.

Once we have fixed the symmetric background dynamics for the collision, the fluctuations behave as a free scalar field with a time- and space-dependent effective frequency.
Using Floquet theory and extending well-developed methods for ODEs to PDEs, we were able to show that the time-dependence of the effective frequency can lead to exponential growth of the symmetry breaking fluctuations.
We also studied the spatial structure of the amplified modes to obtain an understanding of the mechanism responsible for the amplification.
We found generalizations of both broad parametric resonance and narrow parametric resonance.
Due to the spatial dependence of the effective frequency the amplified modes are localized along the collision direction.
For collisions between well defined wall-antiwall pairs, the resulting amplification can be interpreted as Bogoliubov production of particles bound to the walls, i.e., of longitudinally localized modes. The evolution of the effective occupation number shows they are created in bursts when there are well-defined collisions in the background. Once nonlinearity onsets, the strong coupling of the transverse to the longitudinal degrees of freedom breaks this localized particle description.
Numerical lattice solutions are needed to describe the subsequent evolution.

Our detailed study of the unstable modes reveals that the dynamical mechanism responsible for the rapid growth of fluctuations is much more general than our two scalar single-field model examples.
In particular, for collisions between a pair of well defined walls, the most strongly amplified modes are the transverse generalizations of the Goldstone mode resulting from the spontaneous breaking of translation invariance by the domain wall in the symmetry reduced one-dimensional theory.
These modes are present on \emph{any} membrane-like object in a translation invariant theory.
The amplification only relied on a strong deformation of the effective potential binding these fluctuations to the wall.
Such deformations will be extremely common in domain walls formed in scalar field theories, and should also arise in collisions between other membrane-like objects such as D-branes.
Therefore, we expect qualitatively similar results to be obtained in a wide variety of collisions involving membrane-like objects.

The linearized approach taken here cannot tell us what the ultimate fate of the exponentially growing modes will be.
Since the modes have $k_\perp \neq 0$, they do not respect the planar symmetry of the background.
This suggests that this symmetry will be badly broken once the fluctuations become large.
The treatment of the fully nonlinear field evolution is the subject of two companion papers~\cite{ref:bbm2} and~\cite{ref:bbm3}.
We will explore nearly planar symmetric domain wall and nearly SO(2,1) symmetric bubble collisions using high resolution massively parallel scalar field lattice simulations.
These investigations will demonstrate that the nonlinear evolution leads to a complete breakdown of the original symmetry of the background,
including a dissolution of the walls and production of a population of oscillons in the collision region.
This entire process is unique to more than one spatial dimension and is a completely new effect that has not previously been considered in either domain wall or bubble collisions.

\acknowledgments
The authors would like to thank Andrei Frolov and Belle Helen Burgess for useful discussions and comments.
This work was supported by the National Science and Engineering Research Council of Canada and the Canadian Institute for Advanced Research.

\appendix

\section{Collective Coordinate Approximation for Double Well Collisions} \label{sec:collective}
We present a brief derivation of our collective coordinate approximation for the repeated collisions of two walls.
The key step is to drastically reduce the number of degrees of freedom of the system.
We assume the field profile takes the form of an interacting kink-antikink pair
\begin{equation}
  \phi_{bg} = \tanh\left(\gamma\frac{x+r(t)}{\sqrt{2}} \right) - \tanh\left(\gamma\frac{x-r(t)}{\sqrt{2}} \right) - 1, \  \gamma^2 \equiv (1-\dot{r}^2)^{-1} \, . 
  \label{eqn:two_wall_bg}
\end{equation}
This ansatz ignores production of radiation, excitations of the shape mode, and distortion of the kinks due to their mutual interaction.
However, as emphasized in the main text, we wish to separate out the effects of the actual collision from the subsequent evolution and 
 approximation~\eqref{eqn:two_wall_bg} does this.

Our goal is to obtain an effective Lagrangian for the locations, $r$, of the kink and antikink.
This should be of the form for a pair of relativistic point particles interacting through a potential, along with corrections induced by the finite thickness of the kinks.
For simplicity, we assume $\dot{\gamma}=0$.
Terms involving $\dot{\gamma}$ arise only from the kinetic term for the fields, 
and we are dropping a finite thickness correction by ignoring them.
Substituting~\eqref{eqn:two_wall_bg} into the Lagrangian for the field, we find
\begin{align}
  L = &\omega \dot{r}^2\left[S_2(0) + S_2(\omega r)\right]  \notag \\
      &- \omega \left[S_2(0) - S_2(\omega r)\right] \notag \\
      &- \omega^{-1}\left[\sinh^2(2\omega r)S_2(\omega r) - \sinh^3(2\omega r)S_3(\omega r) +\frac{\sinh^4(2\omega r)}{4}S_4(\omega r) \right], \\
      & {\rm where} \ \omega \equiv \frac{\gamma}{\sqrt{2}}, \  S_n(\alpha) \equiv \int \sech^n(x+\alpha)\sech^n(x-\alpha) dx. 
\end{align}
The required integrals are obtained by considering $\int_{\mathcal{C}} f(z) dz$ for ${f(z) = z \mathrm{sech}^n(z+r)\mathrm{sech}^n(z-r)}$ and the contour $\mathcal{C}$ given by $(-\infty,\infty)\cup [\infty,\infty+i\pi]\cup (\infty+i\pi,-\infty+i\pi)\cup [-\infty+i\pi,-\infty]$, with the result 
\begin{align}
  S_2(\alpha)
   &= \frac{4}{\sinh^2(2\alpha)}\left(\frac{2\alpha}{\tanh(2\alpha)}-1\right) \notag \\
  S_3(\alpha)
   &= \frac{2}{\sinh^5(2\alpha)}\left[4\alpha(2+\cosh(4\alpha)) - 3\sinh(4\alpha)\right] \notag \\
  S_4(\alpha) 
  &= \frac{-4}{\sinh^4(2\alpha)}\left[\alpha\coth(2\alpha)(12-20\coth^2(2\alpha)) - \frac{8}{3} + 10\coth^2(2\alpha) \right] \, . 
\end{align}
Note that the interactions between the kink and antikink depend on their relative speeds as well as on their separations.
This is because these interactions are generated by integrals of the overlap between the kink and antikink.
As their speeds increase, the kinks Lorentz contract which changes the amount of overlap.
When the kink and antikink are far apart this overlap is exponentially small.
Therefore, we make an additional approximation:
when performing the integrals we keep overall $\gamma$ multipliers only on the terms with no $r$ dependence.

\begin{figure}[!ht]
  \centering
  \includegraphics[width=0.6\linewidth]{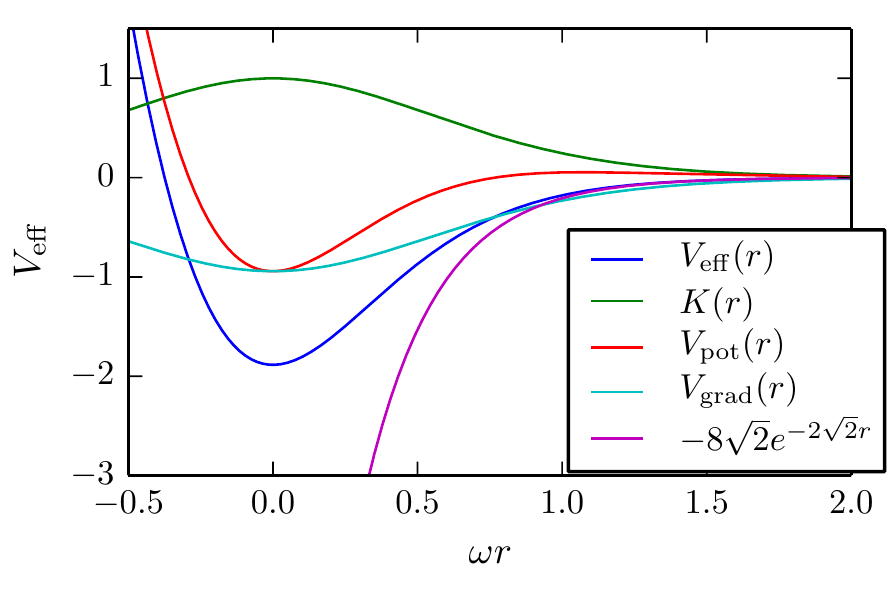}
  \caption[$V_{eff}(r)$ and $K(r)$ for effective single-particle Lagrangian describing kink-antikink separation]{The effective potential $V_{eff}(r)$ and noncanonical contribution to the kinetic term $K(r)$ for our effective single-particle Lagrangian describing the separation of the kink and antikink pair.  Also included are the individual contributions from the gradient energy ($V_{grad}$) and potential energy ($V_{pot}$) in the original scalar field Lagrangian.  For comparison, we have also included the asymptotic potential $V(r) \xrightarrow{\scriptscriptstyle \omega r \gg 1} -8\sqrt{2}e^{-2\sqrt{2}r}$ for $\omega r \gg 1$.}
  \label{fig:r_effective_potential}
\end{figure}

The effective Lagrangian becomes
\begin{equation}
  L[r(t)] = -\frac{4\sqrt{2}}{3}\sqrt{1-\dot{r}^2} - V_{eff}(r) +\gamma K(r)\frac{\dot{r}^2}{2}
  \label{eqn:lagrangian_collective}
\end{equation}
with an the effective potential
\begin{align}
  -V_{eff} \equiv \omega \bigg[ &4\left(1+\gamma^{-2}\right) -16\omega r\gamma^{-2}
    +8\left(-\omega r -3\gamma^{-2} +\omega r\gamma^{-2} \right)\coth(2\omega r) \notag \\  
     &+4\left( -1 + +5\gamma^{-2} + 12\omega r \gamma^{-2} \right)\coth^2(2\omega r) 
    + 8\omega r\left(1 - 5\gamma^{-2}\right)\coth^3(2\omega r) \bigg]
\end{align}
and
\begin{equation}
  K(r) = \frac{S_2(\omega r)}{S_2(0)} \, .
\end{equation}
The multiplier on the relativistic kinetic term can be identified as $2M$ where $M=\frac{2\sqrt{2}}{3}$ is the energy of a single kink at rest.
This potential vanishes exponentially fast for $\omega r \gg 1$.
We only consider bound motions in this paper, so we must have $E - \frac{4\sqrt{2}}{3} < 0$.
Therefore, at large $r$ the walls move nonrelativistically and we can set $\gamma \approx 1$.
This is incorrect for $\omega r \lesssim 1$, but for small separations the kink and antikink are close to each other and interacting strongly.
In the small separation regime, the kink profiles will be deformed, making our ansatz invalid. 
Setting $\gamma=1$,  the resulting effective potential, given by
\begin{equation}
  2^{-1/2}V_{eff} \equiv -4 + 8\omega r + 12\coth(2\omega r) - (24\omega r+8)\coth^2(2\omega r) + 16\omega r\coth^3(2\omega r) \, ,
  \label{eqn:collective_potential}
\end{equation}
is plotted in~\figref{fig:r_effective_potential} along with the noncanonical part of the kinetic coupling.

\begin{figure}
  \centering
  \includegraphics[width=0.6\linewidth]{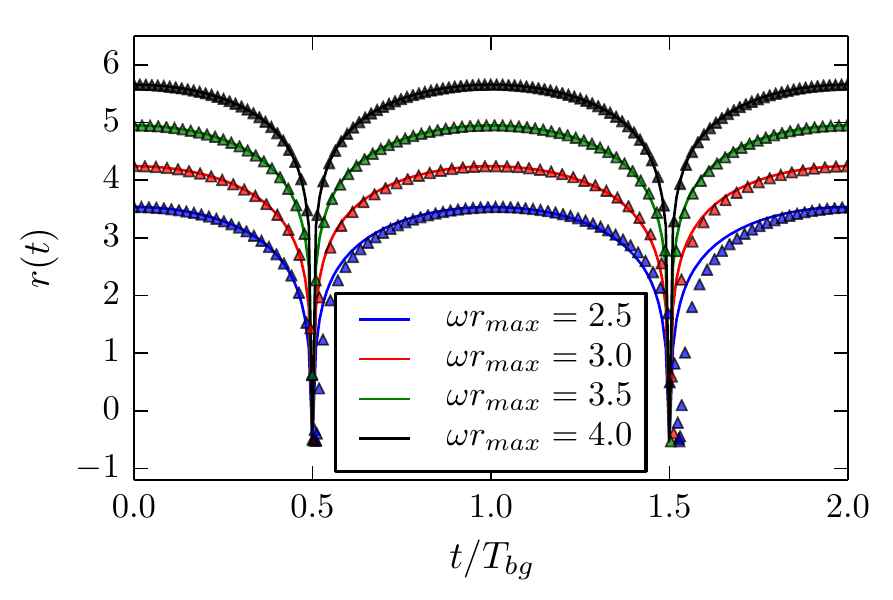}
  \caption[Comparison of analytic approximation to the evolution of $r(t)$ and numerical simulation of the effective equations for the background]{Comparison of the analytic approximation for the evolution of $r(t)$ with numerical simulation of the effective equations for the background.  The solid lines are the analytic approximation~\eqref{eqn:r_background} and the triangles are a numerical solution to the equations of motion for the Lagrangian~\eqref{eqn:lagrangian_collective}.   Thus it is very accurate apart from a small lengthening of the period in the full solution.  The accuracy improves as we increase the initial separation. 
}
  \label{fig:collective_backgrounds}
\end{figure}

Using this effective action we construct an analytic approximation to the background motion.
During most of the motion the walls are well separated with $\omega r \gg 1$.
Therefore, we can approximate the motion as occurring in the potential $V_{eff}(r) \approx -8\sqrt{2}e^{-2\sqrt{2} r}$.
The noncanonical contribution to the kinetic term vanishes exponentially in this limit as well, so we will set it to zero.
Finally, for bound motions we also have $\gamma \approx 1$ so we can approximate the relativistic kinetic term by its nonrelativistic limit.
When $K(r)=0$, energy conservation gives the period $T$ as
\begin{equation}
  t = \sqrt{M} \int_{r_{max}}^r \frac{d\tilde{r}}{\sqrt{V(r_{max})-V(\tilde{r})}} \implies T = 2\sqrt{M}\int_{r_{max}}^{r_{min}} \frac{dr}{\sqrt{V(r_{max})-V(r)}} \, .
\end{equation}
Approximating the full motion by the $\omega r \gg 1$ potential, we find
\begin{equation}
  r(t) = r_{max} + \frac{1}{2\sqrt{2}}\log\left(\cos^2\left(\frac{\pi t}{T}\right)\right)\qquad T = \frac{\pi\sqrt{M}}{\sqrt{2}\sqrt{|V_{eff}(r_{max})|}} = \frac{\pi}{2\sqrt{6}}e^{\sqrt{2}r_{max}}
\end{equation}
Energy conservation implies a minimum value for $r$, but this equation allows $r \to -\infty$. We cure this by cutting off the logarithmic divergence: 
\begin{equation}
  r(t) = r_{max} + \frac{1}{2\sqrt{2}}\log\left(\cos^2\left(\frac{\pi t}{T}\right) + e^{-2\sqrt{2}(r_{max}-r_{min})} \right), 
  \label{eqn:r_background}
\end{equation}
which enforces the condition $r(T/2) = r_{min}$ where $V_{eff}(r_{min}) = V_{eff}(r_{max})$ and $r_{min} < 0$.
In~\figref{fig:collective_backgrounds} we compare the accuracy of this analytic approximation to a full solution of the equations for the Lagrangian~\eqref{eqn:lagrangian_collective}.
We have approximated $\gamma=1$ in $K(r)$ and $V_{eff}(r)$ but have otherwise included both the noncanonical kinetic correction and relativistic corrections.
Our approximation~\eqref{eqn:r_background} is very accurate for most of the evolution.

The above procedure can be generalized to the asymmetric well.
However, the main effect of setting $\delta \neq 0$ is to break the $Z_2$ symmetry so that the two potential minima are no longer degenerate.
This introduces a contribution to the effective potential of form $[V(\phi_f)-V(\phi_t)]r$ for $\omega r \gg 1$, which induces a constant force driving the wall and antiwall towards each other.

\section{Numerical Approach and Convergence Tests}
\label{sec:numerics}
In this appendix we summarize the numerical techniques used in this paper.
For time-evolution in our codes, different schemes were used for the nonlinear background~\eqref{eqn:wall_background} and the linear fluctuations~\eqref{eqn:wall_linearfluc}.
For the one-dimensional nonlinear wave equation~\eqref{eqn:wall_background}, we used a 10th order accurate Gauss-Legendre quadrature based method.  This is a specific choice of an implicit Runge-Kutta process.
Given an initial condition ${\bf y}_t$  to $d{\bf y}/dt = {\bf H}({\bf y})$ at time $t$, the solution at time $t+dt$ is obtained by solving
\begin{align}
  {\bf f}^{(i)} &= {\bf H}\left({\bf y}_t + dt\sum_{j=1}^\nu a_{ij}{\bf f}^{(j)}\right) \\
  {\bf y}_{t+dt} &= {\bf y}_t + dt\sum_{i=1}^\nu b_i{\bf f}^{(i)}
\end{align}
where $a_{ij}$ and $b_{i}$ are numerical constants defining the process.
For Gauss-Legendre methods, $a_{ij}$ and $b_{i}$'s are solutions to
\begin{align}
  \label{eqn:gl_coeffs}
  \notag \sum_{j=1}^\nu a_{ij}c_j^{l-1} &= \frac{c_i^l}{l} \qquad l=1,\dots,\nu \\
  \sum_{j=1}^\nu b_jc_j^{l-1} &= l^{-1} \, .
\end{align}
The $c_i$'s are the roots of $P_\nu(2c-1)$ where $P_\nu(x)$ is the Legendre polynomial of degree $\nu$.
These relations arise by approximating the time-evolution integrals using Gauss-Legendre quadrature.
Because of the excellent convergence properties of quadrature approximations, the result is an order $2\nu$ integrator.\footnote{Here we take the order of the integrator (denoted by n) to be the highest power in $dt$ for which the approximate solution is exact.  This means the leading order error term is $\sim dt^{n+1}$.}
Explicit formulae for $\nu$ up to 5 are given in Table 2 of Butcher~\cite{Butcher:1964}, but it is easier in practice to simply solve~\eqref{eqn:gl_coeffs} numerically.

For the linear fluctuation equation~\eqref{eqn:wall_linearfluc}, we employ Yoshida's~\cite{Yoshida:1990} operator-splitting technique that was introduced into the preheating community by Frolov and Huang.  For further details see, for example,~\cite{Huang:2011gf,Sainio:2012mw,Yoshida:1990}.  For this set of integrators, the solution to $df/dt = H(f)$ is first written as $f(t+dt) = e^{{\bf H}dt}f(t)$, where ${\bf H}$ should now be interpreted as an operator acting on $f$.
We decompose ${\bf H}=\sum_i{\bf H_{i}}$ so the action of each individual ${\bf H_i}$ on $f$ is simple to compute accurately.
Finally, we re-express the time evolution operator $U\equiv e^{{\bf H}dt}$ as a product of exponentials for the individual ${\bf H}_i$ operators, $e^{{\bf H}dt} = U(w_M)U(w_{M-1})\dots U(w_0)U(w_1)\dots U(w_M) + \mathcal{O}(dt^{n+1})$, where $U(w_i) \equiv e^{w_i{\bf H_1} dt/2}e^{w_i{\bf H_2} dt}e^{w_i{\bf H_1}dt/2}$ is a second-order accurate time-evolution operator for time-step $w_idt$ and we have specialized to the case of an operator split into only two parts for simplicity.
Via clever choices of the number and value of the numerical coefficients $w_i$, integrators of various orders $n$ may be constructed.
For this paper, we use an $\mathcal{O}(dt^6)$ method with coefficients $w_i$ given by
\begin{align}
  \notag w_1 &= -1.17767998417887100694641568096431573 \\
  \notag w_2 &= 0.235573213359358133684793182978534602 \\
  \notag w_3 &= 0.784513610477557263819497633866349876 \\
  w_0 &= 1 - 2(w_1+w_2+w_3) = 1.31518632068391121888424972823886251 \, .
\end{align}

The Gauss-Legendre and Yoshida approaches are both symplectic integrators for Hamiltonian systems.
Therefore we find it convenient to use Hamilton's form for the evolution equations.
With the exception of the collective coordinate location for the bouncing walls in the double well, 
all of the Hamiltonians used in this paper can be split into two exactly solvable pieces so that $\mathcal{H}=\mathcal{H}_1+\mathcal{H}_2$.
For reference, we provide these splits below, even though they are not required for the Gauss-Legendre method used to solve the nonlinear background wave equations.
For the planar walls the split terms are (up to an overall normalization)
\begin{equation}
  \mathcal{H}_{planar,1} = \sum_i \frac{\pi_{\phi,i}^2}{2} \, , \qquad \mathcal{H}_{planar,2} = \sum_i\frac{G[\phi_i]}{2dx^2} + V(\phi_i) \, .
  \label{eqn:planar_hamiltonian}
\end{equation}
The linearized fluctuations evolve in the Hamiltonian
\begin{equation}
  \mathcal{H}_{fluc,1} = \sum_i \frac{\pi_{\delta\phi,i}^2}{2} \, , \qquad \mathcal{H}_{fluc,2} = \sum_i \frac{G[\delta\phi_i]}{2dx^2} + \frac{1}{2}V''(\phi_{bg}(x,t))\delta\phi_i^2 \, .
  \label{eqn:fluc_hamiltonian}
\end{equation}
The Hamiltonian for the $SO(2,1)$ invariant bubbles is
\begin{equation}
  \mathcal{H}_{bubbles,1} = \sum_i \frac{\pi_{\phi,i}^2}{2s^2} \, , \qquad \mathcal{H}_{bubbles,2} = \sum_i s^2\left( \frac{G[\phi_i]}{2} + V(\phi_i) \right) \, .
  \label{eqn:bubble_hamiltonian}
\end{equation}
In all three cases, $\pi_{f,i}$ represents the canonical momentum for field $f$ at lattice site $i$.
The operator $G[\phi_i]$ is a discrete approximation to $\left(\partial_x\phi(x_i)\right)^2$.

We now describe the spatial discretization of the system.
For all production runs we used a Fourier pseudospectral approximation for the field derivatives.
The only derivative appearing in the various equations of motion is the one-dimensional \laplacian\ along the collision axis $\partial_{xx}$.
Therefore, in practice the system was evolved in real space, with the \laplacian\ evaluated in Fourier space through the use of the FFT.
Although the resulting FFT and inverse FFT are numerically more expensive than a finite-difference approximation,
the continuum limit is approached much more rapidly as seen in~\figref{fig:floquet_convergence}.
This is especially important when computing Floquet exponents, as our approach requires solving $2N$ individual PDE's in order to form the fundamental matrix where $N$ is the number of grid points.
As well, in order to maintain a fixed accuracy in the time-integration, the ratio $dx/dt$ should be kept constant meaning that the total work required scales as $N^3$ for a finite-difference approximation and $N^3\log(N)$ for a pseudospectral approach.\footnote{When comparing run times, the reader should keep in mind that the limiting factor on modern computing architecture is often the speed at which data can be obtained from memory, not necessarily the number of floating point operations.  However, for the problems we were concerned with the spectral approach proved to be much faster if an accuracy better than the tenth of a percent level was desired.}
Since the pseudospectral calculations converge using much fewer lattice sites than the finite-difference stencils, the pseudospectral approach ends up requiring less CPU time yet is (orders of magnitude) more accurate.

To provide independent verification of our results,
we also performed several runs using finite-difference discretizations of $G[\phi_i]$.
The Hamiltonian was discretized directly, thus ensuring a consistent discretization of $\nabla^2\phi$ and $(\nabla\phi)^2$.
We tested with both second-order accurate
\begin{equation}
 (\nabla\phi)^2dx^2 \approx G[\phi_i]_2 = \frac{1}{2}\left[(\phi_{i+1}-\phi_i)^2 + (\phi_{i-1}-\phi_i)^2\right]
\end{equation}
and fourth-order accurate
\begin{equation}
  (\nabla\phi)^2dx^2 \approx G[\phi_i]_4 = \frac{-1}{24}\left[(\phi_{i+2}-\phi_i)^2 + (\phi_{i-2}-\phi_i)^2\right] + \frac{2}{3}\left[(\phi_{i-1}-\phi_i)^2 + (\phi_{i+1}-\phi_i)^2 \right] 
\end{equation}
finite-difference stencils, where $dx$ is the lattice grid spacing.
The corresponding \laplacian\ stencils $L[\phi_i]/dx^2$ satisfying $\sum_i G[\phi_i] + \phi_i L[\phi_i] = 0$ (on periodic grids) are then the familiar second-order accurate 
\begin{equation}
  \frac{\partial^2\phi}{\partial x^2}dx^2 \approx L[\phi_i]_2 = (\phi_{i+1}-\phi_{i-1}) - 2\phi_i
\end{equation}
and fourth-order accurate
\begin{equation}
  \frac{\partial^2 \phi}{\partial x^2}dx^2 \approx L[\phi_i]_4 = \frac{-1}{12}(\phi_{i-2}+\phi_{i+2}) + \frac{4}{3}(\phi_{i-1}+\phi_{i+1}) + \frac{-5}{2}\phi_i 
\end{equation}
centered differences.

\subsection{Convergence Tests}
The combination of high-order time-integrations and spectrally accurate derivative approximations leads to a rapid convergence of both the nonlinear field evolution used to study the background dynamics and the Floquet exponents obtained by solving the perturbation equations.

\begin{figure}[t]
  \includegraphics[width=0.48\linewidth]{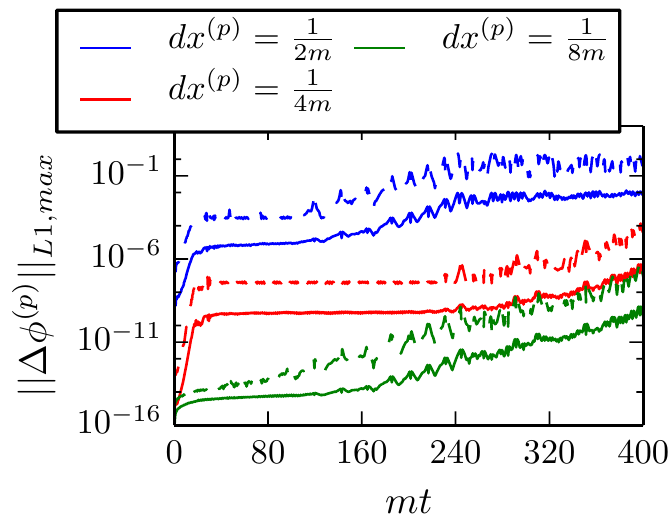}
  \hfill
  \includegraphics[width=0.48\linewidth]{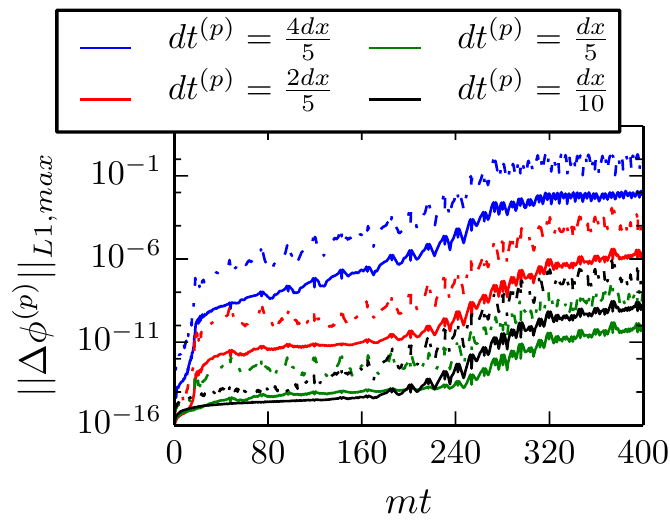} \\
  \includegraphics[width=0.48\linewidth]{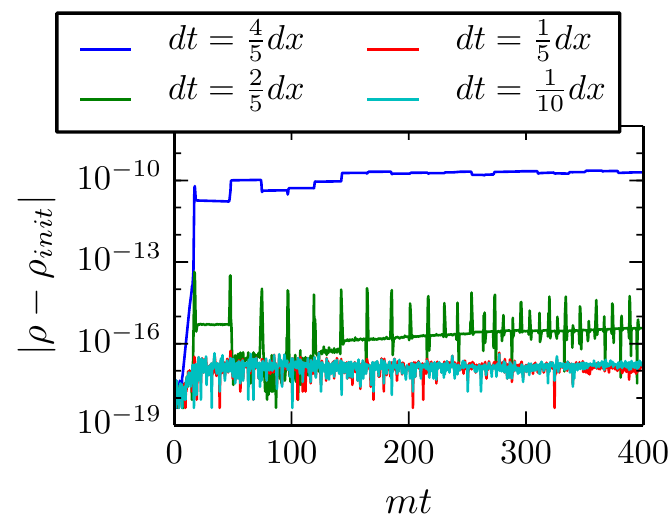}
  \hfill
  \includegraphics[width=0.48\linewidth]{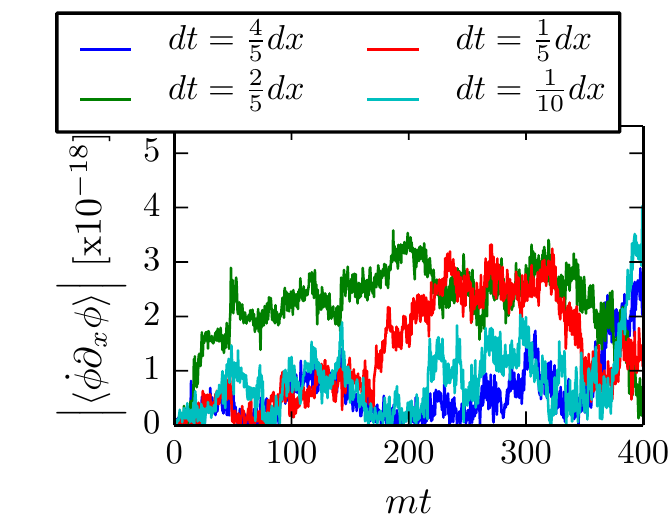}
  \caption[Convergence of the one-dimensional lattice code for the double-well potential]{Convergence of our one-dimensional lattice code for the double-well potential with $\delta=1/30$, $mL=1024$ and various choices of grid spacing $dx$ and time step $dt$.  We plot the two norms defined in~\eqref{eqn:field_norms}.  The accumulating errors at late times are due to small errors in the oscillation frequency and initial phase of the oscillon that has formed at the origin.  Decreasing the time step below $dx/5$ does not lead to a decrease in this error, suggesting that it arises due to machine roundoff.  In the bottom row we demonstrate the convergence of both energy (\emph{bottom left}) and momentum (\emph{bottom right}) of the system for the same choices of $dt$ as in the top right plot.
Since both (conserved) quantities are constant to very high accuracy, this demonstrates that our time-stepping has converged.}
  \label{fig:nonlinear_convergence}
\end{figure}

Several measures of convergence are shown in~\figref{fig:nonlinear_convergence} for the nonlinear wave equation with initial conditions as in the bottom right panel of~\figref{fig:kink_collision}.
In the top row we show the pointwise convergence of the solution as we vary the number of grid points $N$ (or equivalently the grid spacing) and the time-step $dt$,
 thus independently testing our Fourier spatial discretization and our Gauss-Legendre integrator.  
We consider two closely related measures,
\begin{align}
 \notag \lVert \Delta\phi^{(p)}\rVert_{L1} &\equiv N_{base}^{-1}\sum_{\{x_i\}}|\phi^{(p+1)}(x_i)-\phi^{(p)}(x_i)| \\
        \lVert\Delta\phi^{(p)}\rVert_{max}  &\equiv \max_{\{x_i\}}{|\phi^{(p+1)}(x_i)-\phi_i^{(p)}(x_i)|}
 \label{eqn:field_norms}
\end{align}
where $\phi^{(p)}$ denotes the numerical solution for the $pth$ approximation (here either the number of grid points or the time step),
and we take $N^{(p+1)}/N^{(p)} = 2 = dt^{(p)}/dt^{(p+1)}$.
To compare solutions with different spatial resolutions, we took all sums over the grid from the $N=2048$ simulation.
The top left panel, shows that the solution rapidly converges (to the level of machine precision) as we increase the spatial resolution,
exactly as we expect for a properly resolved pseudospectral code.
The top right panel shows that the growing error at late times is due to errors in the time-stepping rather than in the spatial discretization.
This is not really a time-stepping issue, but a demonstration that making a pointwise comparison of the fields is not necessarily the best measure of convergence.
In particular, the errors that accumulate at late time occur because we have an oscillating localized blob of field.
Small errors in the oscillation frequency then lead to errors in the instantaneous value of the field.
These errors manifest themselves as accumulating changes in the pointwise differences at late times.
As a further test of our time evolution we check the conservation of the energy density ${\rho = \langle T^{00}\rangle = \langle\frac{\dot{\phi}^2}{2}+\frac{(\nabla\phi)^2}{2}+V(\phi)\rangle}$ and field momentum ${P^x = \langle T^{0x}\rangle = -\dot{\phi}\partial_x\phi}$ for a range of time steps $dt$.
$T^{\mu\nu}$ is the energy-momentum tensor of $\phi$ and $\langle\cdot\rangle$ denotes a spatial average.
For all choices of time step, the field momentum is conserved to machine precision, 
while for $dt \le dx/5$ the energy is also conserved to machine precision.

\begin{figure}[t]
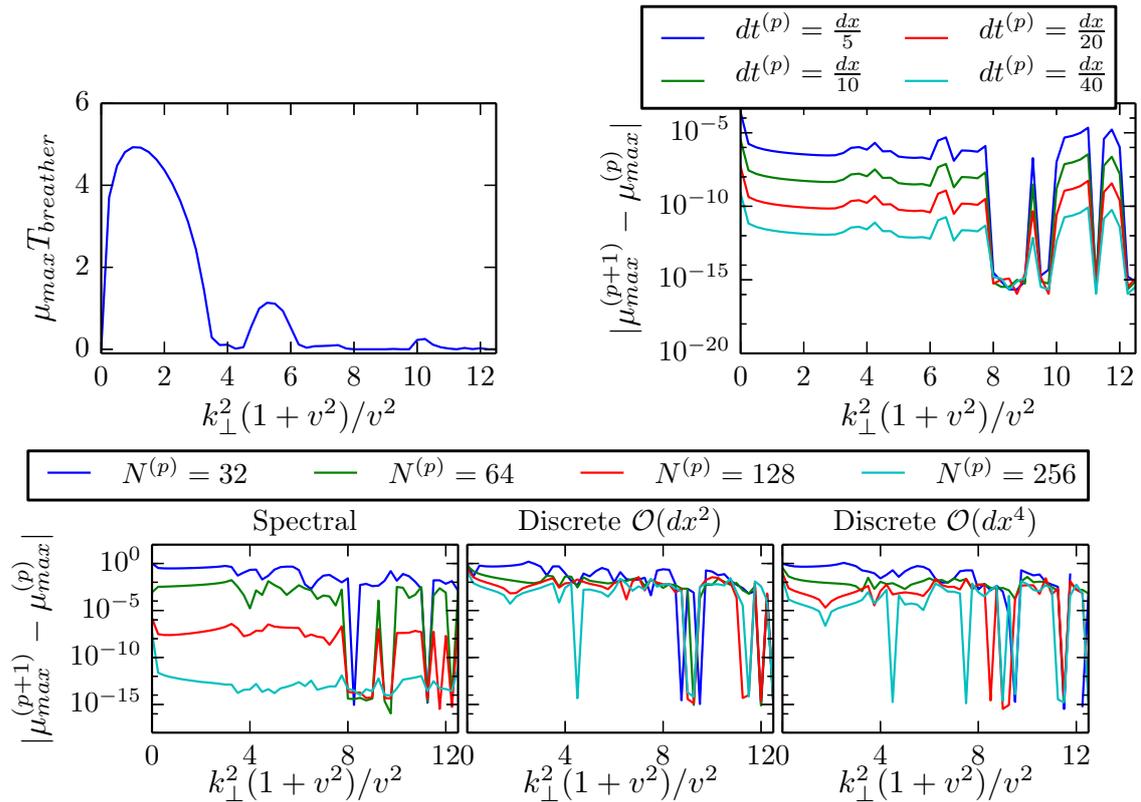

  \includegraphics[width=0.48\linewidth]{{{floquet_v0.5_n64_a20}}}
  \hfill
  \includegraphics[width=0.48\linewidth]{{{evalue_convergence_n64}}} \\
  \includegraphics[width=0.96\linewidth]{{{floquet_convergence_varydx_multipanel}}}
  \caption[Convergence plots for the largest Lyapanov exponent around a sine-Gordon breather]{Convergence of the largest Lyapanov exponent around a sine-Gordon breather with $v=0.5$ for a range of $k_{\perp}^2$ values.  The simulation lattice had length $L=58$.  For reference, the top left panel shows $\mu_{\mathrm{max}}T_{\mathrm{breather}}$ with $N=64$ and $dt=dx/20$.
These are the same parameters used along the line $v=0.5$ in the instability chart in~\figref{fig:sg_chart}.  
In the top right panel, we show the difference in the numerically determined values of $\mu_{\mathrm{max}}T_{\mathrm{breather}}$ holding the number of grid points fixed (at $N=64$) while varying the discrete time step $dt$.
A pseudospectral derivative approximation was used for each run and a sixth-order Yoshida integrator for the time evolution.  
The bottom row shows the convergence properties as the grid spacing is decreased, using a pseudospectral (\emph{bottom left}), second-order finite-difference (\emph{bottom centre}) and fourth-order finite-difference (\emph{bottom right}) approximation for the \laplacian.  In all three panels, we took $dx/dt = 20$ and used a sixth-order accurate Yoshida scheme.  
Because of the rapid convergence, the instability charts (with the exception of the $N=32$ case) are visually indistinguishable from the top left panel.}
  \label{fig:floquet_convergence}
\end{figure}

To understand the accuracy of our Floquet exponents, and to demonstrate the great gains in accuracy obtained via a pseudospectral approach relative to a finite-difference scheme, we show convergence plots for the maximal Floquet exponent in~\figref{fig:floquet_convergence}.
Here we can directly compare the individual Floquet exponents, so we plot
\begin{equation}
  \Delta\mu^{(p)} \equiv |\mu_{max}^{(p+1)}-\mu_{max}^{(p)}| \, .
\end{equation}
For orientation, the top left panel shows $\mu_{max}T_{breather}$ for the choice $v=0.5$ and a range of $k_{\perp}$ values.
The remaining panels show the convergence properties of the exponents in the top left panel for various numerical schemes.
The top right panel shows the convergence rate as the time-step is varied.
As expected for a sixth-order accurate integrator, the error decreases rapidly, although not quite as quickly as for the Gauss-Legendre integrator.
Also of note is that the error decreases uniformly for all values of $k_\perp$ (except for those values that are already at machine roundoff levels)
indicating that important features such as the locations of stability bands where $\mu_{max}=0$ are not shifting around as the time-step is varied.
In the bottom row we show similar convergence plots as the number of grid points are varied while holding the simulation box size fixed.
We show results for a pseudospectral approximation, a second-order accurate finite-difference stencil, and a fourth-order accurate finite-difference stencil.
As promised, the psuedospectral method converges much more rapidly than the finite-differencing methods.
Also of note is the uniform convergence for all $k_\perp$ with the pseudospectral approximation, whereas the convergence is non-uniform for the finite-difference methods.
Taking the far right 4th order chart as an example, there are several extreme spikes in the region $k_\perp^2(1+v^{-2})$ for which the difference between the $N=128$ and $N=256$ approximation is of order machine precision, but then rises to the $10^{-3}$ level when comparing to the $N=512$ solution.
The ultimate source of these appearing and disappearing spikes is a slight shifting of the edges of the stability bands as the resolution is varied.

\bibliography{linear_fluc_full}{99}

\end{document}